\documentclass[prd,draft,amsfonts,amssymb,amsmath,showpacs]{revtex4}

\begin{document}

\title{Supersymmetric Nambu$-$Jona-Lasinio Model
on ${\cal N}=1/2$ four-dimensional Non(anti)commutative Superspace}
\author{Tadafumi Ohsaku}
\affiliation{Institut f\"{u}r Theoretische Physik, Universit\"{a}t zu K\"{o}ln, 50937 K\"{o}ln, Deutschland}

\date{\today}


\newcommand{\bmx}{\mbox{\boldmath $x$}}
\newcommand{\bmy}{\mbox{\boldmath $y$}}
\newcommand{\bmk}{\mbox{\boldmath $k$}}
\newcommand{\bmp}{\mbox{\boldmath $p$}}
\newcommand{\bmq}{\mbox{\boldmath $q$}}
\newcommand{\bmP}{\mbox{\boldmath $P$}}  
\newcommand{\kfey}{\ooalign{\hfil/\hfil\crcr$k$}}
\newcommand{\pfey}{\ooalign{\hfil/\hfil\crcr$p$}}
\newcommand{\qfey}{\ooalign{\hfil/\hfil\crcr$q$}}
\newcommand{\Deltafey}{\ooalign{\hfil/\hfil\crcr$\Delta$}}
\newcommand{\nablafey}{\ooalign{\hfil/\hfil\crcr$\nabla$}}
\newcommand{\Dfey}{\ooalign{\hfil/\hfil\crcr$D$}}
\newcommand{\partfey}{\ooalign{\hfil/\hfil\crcr$\partial$}}
\def\sech{\mathop{\rm sech}\nolimits}

\begin{abstract}

We construct the Lagrangian of the ${\cal N}=1$ four-dimensional 
generalized supersymmetric Nambu$-$Jona-Lasinio ( SNJL ) model,
which has ${\cal N}=1/2$ supersymmetry ( SUSY ) on non(anti)commutative superspace.
A special attention is paid to the examination on the nonperturbative quantum dynamics:
The phenomenon of dynamical-symmetry-breaking/mass-generation on the deformed superspace is investigated.
The model Lagrangian and the method of SUSY auxiliary fields of composites are examined
in terms of component fields.
We derive the effective action, examine it, and solve the gap equation for self-consistent mass parameters.
( Keywords: Superspaces, Non-Commutative Geometry, Supersymmetric Effective Theories. )

\end{abstract}


\maketitle

\section{Introduction}

Field theories on noncommutative spacetime defined by the Heisenberg algebra
$[x^{\mu},x^{\nu}]=i\Theta^{\mu\nu}\ne 0$ have various interesting properties~[1,2],
and considered that they might capture important aspects of quantum gravity.
Recently, it was shown that, low-energy effective descriptions of four-dimensional superstrings 
in the self-dual graviphoton background fields $F^{\alpha\beta}$ 
( with setting anti-self-dual part $F^{\dot{\alpha}\dot{\beta}}=0$ )
give the deformations of superspaces~[3,4] ( in a ten-dimensional case, see Ref.~[5,6] )
as ${\cal N}=1\to{\cal N}=1/2$, with giving
\begin{eqnarray}
\{\theta^{\alpha},\theta^{\beta}\} &=& 2\alpha'F^{\alpha\beta}.
\end{eqnarray}
Here, $\alpha'$ is the inverse of a string tension.
Motivated by these examinations, various physical and mathematical aspects of the deformation of 
superspace obtained much attentions~[7-22].
Under this context, supersymmetric ( SUSY ) Euclidean field theories on non(anti)commutative superspace 
were first considered in Ref.~[7].
The ${\cal N}=1/2$ deformed Wess-Zumino ( WZ ) model~[7,11,12,21] and 
deformed SUSY gauge models with(out) matter~[7,10,13,14] were examined by several works, 
and the preservation of the locality, Lorentz symmetry
and the renormalizabilities of all orders in perturbative expansions in these models were proved~[13]. 
It was also shown in Refs.~[11,12] that the chiral superpotential of the WZ model obtains 
quantum radiative corrections while the antichiral superpotential part of it 
obeys the nonrenormalization theorem by the perturbative analyses.
By the quantum dynamics of itself, the WZ model gets a term which is not included
in the classical deformed action of it~[11,12,21].
The methods to construct generic sigma models both on ${\cal N}=1$ four-dimensional 
and ${\cal N}=2$ two-dimensional deformed superspaces were proposed, 
and the component field expressions for them were derived~[15-19].
Until now, various examinations were performed under the classical or perturbative quantum levels. 
Hence, it is interesting for us to examine nonperturbative effects of quantum dynamics of a model
on a deformed superspace.

In this paper, we investigate a nonperturbative quantum dynamics in a SUSY model 
on the ${\cal N}=1/2$ four-dimensional non(anti)commutative superspace.
Of particular interest in here is the phenomenon of dynamical symmetry breaking
of the mass generation of the BCS-NJL mechanism ( Bardeen, Cooper, Schrieffer, Nambu and Jona-Lasinio )~[23,24,25]. 
Until now, there is no attempt for an investigation of the nonperturbative dynamical mass generation
in literature.
We introduce an ${\cal N}=1$ four-dimensional supersymmetric Nambu$-$Jona-Lasinio ( SNJL ) model~[26-31], 
regarded as the simplest example which might realize the dynamical symmetry-breaking/mass-generation,
and examine the characteristic aspects of the model on the deformed superspace. 
The nonrenormalization theorem in the ${\cal N}=1$ case forbids the
dynamical break down of chiral symmetry by the mass generation, 
while it can take place when ${\cal N}=1$ SUSY is completely broken by introducing
a non-holomorphic soft-SUSY-breaking mass~[26-31].
Because the nonrenormalization theorem is partly broken in the ${\cal N}=1/2$ superspace, 
it is interesting for us whether the SNJL model gets a dynamical mass radiatively or not.
In Sec. II, we summarize the definitions and algebraic properties of the deformed superspace.
The SNJL model in the deformed superspace will be examined in detail in Sec.III.
The method of SUSY auxiliary fields of composites on the deformed superspace is investigated.
The one-loop effective action in the leading order of the large-$N$ expansion 
will be evaluated by the superspace formalism, 
and the gap equation for a self-consistent mass will be derived and solved.
The comments on further investigations for our method are given in Sec. IV.

\section{Superfields on Non(anti)commutative Superspace}

First, we summarize several algebraic definitions of ${\cal N}=1/2$ four-dimensional 
non(anti)commutative superspace for the preparation of our calculations given later.
The notations and spinor conventions follow both Ref.~[7] and the textbook of Wess and Bagger~[32].
The (anti)commutation relations for bosonic coordinates $x^{\mu}$ and fermionic coordinates 
$\theta^{\alpha}$, $\bar{\theta}^{\dot{\alpha}}$ ( $\alpha, \dot{\alpha} = 1, 2$ ) 
are required to satisfy the following relations:
\begin{eqnarray}
\{\theta^{\alpha},\theta^{\beta}\} &\equiv& C^{\alpha\beta} = C^{\beta\alpha} \ne 0, \\
\{\bar{\theta}^{\dot{\alpha}},\bar{\theta}^{\dot{\beta}}\} &=& 
\{\bar{\theta}^{\dot{\alpha}},\theta^{\alpha}\}= [\bar{\theta}^{\dot{\alpha}},x^{\mu}] =0.
\end{eqnarray}
$C^{\alpha\beta}$ are the deformation parameters of superspace.
Hence $(\theta^{\alpha})^{*}\ne\bar{\theta}^{\dot{\alpha}}$, 
they are independent with each other. 
At this stage, the Lorentz symmetry $SO(4)=SU(2)_{L}\times SU(2)_{R}$ is explicitly broken to $SU(2)_{R}$ at $C^{\alpha\beta}\ne 0$. 
Due to this fact, we have to work in the Euclidean spacetime ${\bf R}^{4}$,
though the Lorentzian signature $\eta^{\mu\nu}={\rm diag}(-1,1,1,1)$ is used throughout this paper.
Any functions of $\theta$ have to be ordered, and we use Weyl ordering: 
$(\theta^{\alpha}\theta^{\beta})_{W}\equiv\frac{1}{2}[\theta^{\alpha},\theta^{\beta}]
=-\frac{1}{2}\epsilon^{\alpha\beta}\theta\theta$.
For example, $\theta\theta$ is Weyl ordered.
The anticommutator (2) of the Clifford algebra gives nontrivial results 
compared with the ordinary ${\cal N}=1$ SUSY case.
Clearly, the Hermiticity of a theory was lost, and the unitarity might be violated in a field theory 
on non(anti)commutative superspace.
Thus, a potential energy could become complex and it would give an unstable vacuum of a theory.
However, it was proved that the vacuum energy vanishes in the deformed WZ model
because the notion of anti-holomorphicity remains in the part of its antichiral superpotential,
and this fact makes its vacuum stable and vanishes the vacuum energy~[11,12].
This situation might be true in a model which has an antichiral superpotential.
From these exotic characters, it is particularly interesting for us to generalize several methods 
of quantum field theory to cases on the deformed superspace.

The chiral coordinates $y^{\mu}$ are defined as
\begin{eqnarray}
y^{\mu} &\equiv& x^{\mu}+i\theta^{\alpha}\sigma^{\mu}_{\alpha\dot{\alpha}}\bar{\theta}^{\dot{\alpha}},
\end{eqnarray}
while, the antichiral coordinates $\bar{y}^{\mu}$ are given as follows:
\begin{eqnarray}
\bar{y}^{\mu} &\equiv& y^{\mu}-2i\theta^{\alpha}\sigma^{\mu}_{\alpha\dot{\alpha}}\bar{\theta}^{\dot{\alpha}}= x^{\mu}-i\theta^{\alpha}\sigma^{\mu}_{\alpha\dot{\alpha}}\bar{\theta}^{\dot{\alpha}}.
\end{eqnarray}
Several commutators are summarized below:
\begin{eqnarray}
&& [y^{\mu},y^{\nu}] = [y^{\mu},\bar{y}^{\nu}] = [y^{\mu},\theta^{\alpha}] = [y^{\mu}, \bar{\theta}^{\dot{\alpha}}] = 0, \quad [\bar{y}^{\mu},\bar{y}^{\nu}] = 4\bar{\theta}\bar{\theta}C^{\mu\nu},  \nonumber \\
&& [x^{\mu},\theta^{\alpha}] = iC^{\alpha\beta}\sigma^{\mu}_{\beta\dot{\alpha}}\bar{\theta}^{\dot{\alpha}}, \quad [x^{\mu},x^{\nu}] = \bar{\theta}\bar{\theta}C^{\mu\nu}, \quad C^{\mu\nu} \equiv C^{\alpha\beta}\epsilon_{\beta\gamma}(\sigma^{\mu\nu})^{\gamma}_{\alpha} = -C^{\nu\mu}. 
\end{eqnarray}
These commutators make the structure of the deformed superspace more clearly.
Due to the commutation relations $[y^{\mu},y^{\nu}]=0$, 
the supercovariant derivatives can be chosen so as to satisfy the Leibniz rule, 
then it is possible for us to define a chiral superfield on the deformed superspace~[7].
As a consequence, the deformed superspace will be parametrized 
by a set $(y,\theta,\bar{\theta})$ as coordinates in the chiral basis.
We have to keep this fact in mind to perform any computations of quantities given by (anti)chiral
superfields on the deformed superspace.
The supercovariant derivatives and the supercharges are defined by
\begin{eqnarray}
D_{\alpha} \equiv \frac{\partial}{\partial\theta^{\alpha}}+2i\sigma^{\mu}_{\alpha\dot{\alpha}}\bar{\theta}^{\dot{\alpha}}\frac{\partial}{\partial y^{\mu}}, \qquad \overline{D}_{\dot{\alpha}} \equiv -\frac{\partial}{\partial\bar{\theta}^{\dot{\alpha}}},
\end{eqnarray}
and
\begin{eqnarray}
Q_{\alpha} \equiv \frac{\partial}{\partial\theta^{\alpha}}, \qquad 
\overline{Q}_{\dot{\alpha}} \equiv -\frac{\partial}{\partial\bar{\theta}^{\dot{\alpha}}}+2i\theta^{\alpha}\sigma^{\mu}_{\alpha\dot{\alpha}}\frac{\partial}{\partial y^{\mu}},
\end{eqnarray}
respectively. Clearly, they satisfy
\begin{eqnarray}
&0& = \{ D_{\alpha}, D_{\beta} \} = \{ \overline{D}_{\dot{\alpha}}, \overline{D}_{\dot{\beta}} \} = \{ D_{\alpha}, Q_{\beta} \} = \{ \overline{D}_{\dot{\alpha}}, Q_{\beta} \} = \{ D_{\alpha},\overline{Q}_{\dot{\beta}} \} = \{ \overline{D}_{\dot{\alpha}},\overline{Q}_{\dot{\beta}} \} = \{ Q_{\alpha}, Q_{\beta} \}, \nonumber \\
& & \{ \overline{D}_{\dot{\alpha}}, D_{\alpha} \} = -2i\sigma^{\mu}_{\alpha\dot{\alpha}}\frac{\partial}{\partial y^{\mu}}, \quad \{ \overline{Q}_{\dot{\alpha}}, Q_{\alpha} \} = 2i\sigma^{\mu}_{\alpha\dot{\alpha}}\frac{\partial}{\partial y^{\mu}}, \quad \{ \overline{Q}_{ \dot{\alpha} }, \overline{Q}_{ \dot{\beta} } \} = -4C^{\alpha\beta}\sigma^{\mu}_{ \alpha\dot{\alpha} }\sigma^{\nu}_{ \beta\dot{\beta} }\frac{ \partial^{2} }{ \partial y^{\mu}\partial y^{\nu} }. 
\end{eqnarray}

Any products of Weyl-ordered functions of $\theta$ have to be Weyl re-ordered,
and it is implemented by the star product:
\begin{eqnarray}
f(y,\theta)\star g(y,\theta) &\equiv& f(y,\theta)
\exp\Bigl( -\frac{C^{\alpha\beta}}{2}\frac{\overleftarrow{\partial}}{\partial\theta^{\alpha}}\frac{\overrightarrow{\partial}}{\partial\theta^{\beta}} \Bigr)g(y,\theta)  \nonumber \\
&=& f(y,\theta)g(y,\theta)+(-1)^{{\rm deg}f}\frac{C^{\alpha\beta}}{2}\frac{\partial}{\partial\theta^{\alpha}}f(y,\theta)\frac{\partial}{\partial\theta^{\beta}}g(y,\theta) 
-\det C\frac{\partial}{\partial\theta\theta}f(y,\theta)\frac{\partial}{\partial\theta\theta}g(y,\theta),
\end{eqnarray}
where,
\begin{eqnarray}
\frac{\partial}{\partial\theta\theta} &\equiv& \frac{\epsilon^{\alpha\beta}}{4}\frac{\partial}{\partial\theta^{\alpha}}\frac{\partial}{\partial\theta^{\beta}} = Q^{2}, \quad \det C = \frac{1}{4}C^{\mu\nu}C_{\mu\nu}. 
\end{eqnarray}
The derivatives $\frac{\partial}{\partial \theta^{\alpha}}$ should apply at fixed $y^{\mu}$.
From the definitions given above, one finds
\begin{eqnarray}
\theta^{\alpha}\star\theta^{\beta} &=& -\frac{\epsilon^{\alpha\beta}}{2}\theta\theta + \frac{C^{\alpha\beta}}{2}, \quad \theta^{\alpha}\star\theta\theta = C^{\alpha\beta}\theta_{\beta}, \quad \theta\theta\star\theta^{\alpha} = -C^{\alpha\beta}\theta_{\beta}, \nonumber \\
\theta\theta\star\theta\theta &=& -\det C, \quad \theta\sigma^{\mu}\bar{\theta}\star\theta\sigma^{\nu}\bar{\theta} = -\frac{1}{2}\eta^{\mu\nu}\theta\theta\bar{\theta}\bar{\theta}-\frac{1}{2}C^{\mu\nu}\bar{\theta}\bar{\theta}.  
\end{eqnarray}
Because of the commutator $[\bar{y}^{\mu},\bar{y}^{\nu}]=4\bar{\theta}\bar{\theta}C^{\mu\nu} \ne 0$,
the definition of the star product (10) is applicable 
when $f$ and $g$ are given as functions of only $y$, $\theta$ and $\bar{\theta}$.
For a star product of $\bar{f}(\bar{y},\bar{\theta})$ and $\bar{g}(\bar{y},\bar{\theta})$, 
the following re-expression,
\begin{eqnarray}
\bar{f}(\bar{y},\bar{\theta})\star\bar{g}(\bar{y},\bar{\theta}) = 
\bar{f}(\bar{y},\bar{\theta})\exp\Bigl(2\bar{\theta}\bar{\theta}C^{\mu\nu}\frac{\overleftarrow{\partial}}{\partial\bar{y}^{\mu}}\frac{\overrightarrow{\partial}}{\partial\bar{y}^{\nu}} \Bigr)\bar{g}(\bar{y},\bar{\theta})
\end{eqnarray} 
is also useful.

Because $\overline{Q}_{\dot{\alpha}}$ have $\theta^{\alpha}$, they do not commute with the star product:
$\overline{Q}_{\dot{\alpha}}$ are broken generators. In other words, the "translation" symmetry 
of $\bar{\xi}^{\dot{\alpha}}$-directions are explicitly broken,
and we must choose the point of the direction for constructing a field theory. 
On the other hand, $Q_{\alpha}$ are not broken generators, hence the deformed superspace is called as ${\cal N}=1/2$ superspace.
The definition of a chiral superfield $\Phi$ is 
\begin{eqnarray}
\Phi(y,\theta) \equiv \phi(y)+\sqrt{2}\theta\psi(y)+\theta\theta F(y), \qquad
\overline{D}_{\dot{\alpha}}\Phi = 0.
\end{eqnarray}
Hence $\Phi(y,\theta)$ has the Weyl-ordered form.
An antichiral superfield $\bar{\Phi}$ is defined as follows:
\begin{eqnarray}
\bar{\Phi}(\bar{y},\bar{\theta}) \equiv \bar{\phi}(\bar{y})+\sqrt{2}\bar{\theta}\bar{\psi}(\bar{y})+\bar{\theta}\bar{\theta}\bar{F}(\bar{y}), \qquad D_{\alpha}\bar{\Phi} = 0.
\end{eqnarray}
Because $\bar{\Phi}(\bar{y},\bar{\theta})$ includes $\bar{y}$, it has to be Weyl ordered.
It will be written in the chiral coordinates $y^{\mu}$ under the Weyl-ordered form as a function of $\theta$: 
\begin{eqnarray}
\bar{\Phi}(\bar{y},\bar{\theta}) = \bar{\Phi}(y-2i\theta\sigma\bar{\theta},\bar{\theta}) 
= \bar{\phi}(y)+\sqrt{2}\bar{\theta}\bar{\psi}(y)+\bar{\theta}\bar{\theta}\bar{F}(y) + \sqrt{2}\theta\{ i\sigma^{\mu}\partial_{\mu}\bar{\psi}(y)\bar{\theta}\bar{\theta}-i\sqrt{2}\sigma^{\mu}\bar{\theta}\partial_{\mu}\bar{\phi}(y)\}+\theta\theta\bar{\theta}\bar{\theta}\Box\bar{\phi}(y).
\end{eqnarray}
On the deformed superspace, $\Phi$ and $\bar{\Phi}$ are independent with each other:
$(\Phi)^{\dagger}\ne \bar{\Phi}$.
From the preservation of chiralities in both a chiral and an antichiral fields on non(anti)commutative superspace,
we should utilize the chiral projectors for a variation of (anti)chiral superfields,
in perturbative and nonperturbative calculations, so forth, as same as the ordinary ${\cal N}=1$ case.

A star product of (anti)chiral superfields is again a (anti)chiral superfield. 
For example,
\begin{eqnarray}
\Phi_{1}(y,\theta)\star\Phi_{2}(y,\theta) &=& \Phi_{1}(y,\theta)\Phi_{2}(y,\theta)
-C^{\alpha\beta}(\psi_{1})_{\alpha}(y)(\psi_{2})_{\beta}(y) -\det C F_{1}(y)F_{2}(y) \nonumber \\
& & +\sqrt{2}C^{\alpha\beta}\theta_{\beta}[(\psi_{1})_{\alpha}(y)F_{2}(y)-(\psi_{2})_{\alpha}(y)F_{1}(y)], \\
\bar{\Phi}_{1}(\bar{y},\bar{\theta})\star\bar{\Phi}_{2}(\bar{y},\bar{\theta}) &=& 
\bar{\Phi}_{1}(\bar{y},\bar{\theta})\bar{\Phi}_{2}(\bar{y},\bar{\theta})+2\bar{\theta}\bar{\theta}C^{\mu\nu}\frac{\partial}{\partial \bar{y}^{\mu}}\bar{\Phi}_{1}(\bar{y},\bar{\theta})\frac{\partial}{\partial \bar{y}^{\nu}}\bar{\Phi}_{2}(\bar{y},\bar{\theta}), \\
\bar{\Phi}_{1}(\bar{y},\bar{\theta})\star\Phi_{2}(y,\theta) &=& \bar{\Phi}_{1}(y-2i\theta\sigma\bar{\theta},\bar{\theta})\star\Phi_{2}(y,\theta) = \bar{\Phi}_{1}(y-2i\theta\sigma\bar{\theta},\bar{\theta})
\exp\Bigl( -\frac{C^{\alpha\beta}}{2}\frac{\overleftarrow{\partial}}{\partial\theta^{\alpha}}\frac{\overrightarrow{\partial}}{\partial\theta^{\beta}} \Bigr)
\Phi_{2}(y,\theta).
\end{eqnarray}
Any products of (anti)chiral superfields do not commute by the non(anti)commutativity, 
$\Phi_{1}\star\Phi_{2}\ne\Phi_{2}\star\Phi_{1}$, etc.,
there are ambiguities to construct a Lagrangian in terms of products of chiral superfields.
In this paper, we will use the symmetrized star products between (anti)chiral superfields themselves
defined as follows~[15,16]:
\begin{eqnarray}
\Phi_{1}\star\Phi_{2}\star\cdots\star\Phi_{n}|_{sym} &\equiv& \frac{1}{n!}(\Phi_{1}\star\Phi_{2}\star\cdots\star\Phi_{n}+{\rm permutations}),   \\
\bar{\Phi}_{1}\star\bar{\Phi}_{2}\star\cdots\star\bar{\Phi}_{n}|_{sym} &\equiv& \frac{1}{n!}(\bar{\Phi}_{1}\star\bar{\Phi}_{2}\star\cdots\star\bar{\Phi}_{n}+{\rm permutations}),  
\end{eqnarray}
and then
\begin{eqnarray}
\Phi_{1}\star\cdots\star\Phi_{i}|_{sym}\star\bar{\Phi}_{i+1}\star\cdots\star\bar{\Phi}_{n}|_{sym} &\equiv& 
\frac{1}{i!(n-i)!}(\Phi_{1}\star\cdots\star\Phi_{i}+{\rm permutations})   \nonumber \\
& & \times \star(\bar{\Phi}_{i+1}\star\cdots\star\bar{\Phi}_{n}+{\rm permutations}).
\end{eqnarray}
These symmetrized products could be interpreted as a kind of ordering in terms of (anti)chiral superfields.
There are a few choices of symmetrization, namely, taking the symmetrization between a chiral and an antichiral
parts of a product or not, and in general the difference of choices gives different results
of their component field expressions~[16], though at least in the model Lagrangian we will consider in this paper,
there is only a difference of a numerical factor.
For example,
\begin{eqnarray}
\Phi_{1}\star\Phi_{2}|_{sym} &\equiv& \frac{1}{2}(\Phi_{1}\star\Phi_{2}+\Phi_{2}\star\Phi_{1}) = \Phi_{1}\Phi_{2}-\det C\frac{\partial}{\partial\theta\theta}\Phi_{1}\frac{\partial}{\partial\theta\theta}\Phi_{2} = \Phi_{1}\Phi_{2}-\det C(Q^{2}\Phi_{1})(Q^{2}\Phi_{2}),  \\
\bar{\Phi}_{1}\star\bar{\Phi}_{2}|_{sym} &\equiv& \frac{1}{2}(\bar{\Phi}_{1}\star\bar{\Phi}_{2}+\bar{\Phi}_{2}\star\bar{\Phi}_{1}) = \bar{\Phi}_{1}\bar{\Phi}_{2}.
\end{eqnarray}
Hence, the deformation will modify an action functional of a theory through a product of chiral multiplets.
There is no contribution coming from the deformation in a symmetrized product of antichiral superfields, 
because $C^{\mu\nu}=-C^{\nu\mu}$. 
Fortunately, such Lorentz-symmetry breaking parameters are explicitly
removed from a theory under the symmetrization in the superfield level, 
and a theory includes the deformation parameters only in the Lorentz-scalar form $\det C$,
in spite of the Lorentz-symmetry breaking relation defined in Eq. (2).

\section{The Generalized Supersymmetric Nambu$-$Jona-Lasinio Model}

\subsection{Symmetries and Structure}

We employ the following $U(N)_{L}\times U(N)_{R}$-invariant Lagrangian of an ${\cal N}=1$ generalized SNJL model on the four-dimensional deformed superspace:
\begin{eqnarray}
{\cal L}(C) &\equiv& \Bigl[ K(\bar{\Phi}_{\pm},\Phi_{\pm}) \Bigr]_{\theta\theta\bar{\theta}\bar{\theta}} = {\cal L}_{0}+{\cal L}^{(1)}_{I}(C)+{\cal L}^{(2)}_{I}(C), \\
{\cal L}_{0} &\equiv& \Bigl[ \bar{\Phi}_{+}\star\Phi_{+} + \bar{\Phi}_{-}\star\Phi_{-} \Bigr]_{\theta\theta\bar{\theta}\bar{\theta}} 
= \Bigl[ \bar{\Phi}_{+}\Phi_{+}+\bar{\Phi}_{-}\Phi_{-} \Bigr]_{\theta\theta\bar{\theta}\bar{\theta}}, \\
{\cal L}^{(1)}_{I}(C) &\equiv&  \frac{G_{1}}{N}\Bigl[ \bar{\Phi}_{+}\star\bar{\Phi}_{-}|_{sym}\star\Phi_{+}\star\Phi_{-}|_{sym} \Bigr]_{\theta\theta\bar{\theta}\bar{\theta}},  \nonumber \\
&=& \frac{G_{1}}{N}\Bigl[ \bar{\Phi}_{+}\bar{\Phi}_{-}\Bigl(\Phi_{+}\Phi_{-} -\det C(Q^{2}\Phi_{+})(Q^{2}\Phi_{-})\Bigr)\Bigr]_{\theta\theta\bar{\theta}\bar{\theta}}, \\
{\cal L}^{(2)}_{I}(C) &\equiv& \frac{G_{2}}{4N}\Bigl[ 
(\bar{\Phi}_{+}\star\bar{\Phi}_{+}+\bar{\Phi}_{-}\star\bar{\Phi}_{-})\star(\Phi_{+}\star\Phi_{+}+\Phi_{-}\star\Phi_{-}) \Bigr]_{\theta\theta\bar{\theta}\bar{\theta}},  \nonumber \\
&=& \frac{G_{2}}{4N}\Bigl[ (\bar{\Phi}_{+}\bar{\Phi}_{+}+\bar{\Phi}_{-}\bar{\Phi}_{-})\Bigl\{ \Phi_{+}\Phi_{+}+\Phi_{-}\Phi_{-}-\det C(Q^{2}\Phi_{+})^{2}-\det C(Q^{2}\Phi_{-})^{2} \Bigr\}\Bigr]_{\theta\theta\bar{\theta}\bar{\theta}}.  
\end{eqnarray}
Here, $K(\bar{\Phi}_{\pm},\Phi_{\pm})$ is the K\"{a}hler potential of our model, $N$ is a number of flavor.
$G_{1}$ and $G_{2}$ are coupling constants ( $G_{1},G_{2} > 0$ ) they have mass dimension $[{\rm Mass}]^{-2}$, thus
this model is nonrenormalizable under a power-counting analysis in the ${\cal N}=1$ case.
In the observation of the WZ model, the operators coming from the deformation carry mass dimensions
larger than four, though they obtain radiative corrections with divergences at most logarithmically~[11,12].
This fact is one of the bases of the proof of renormalizability of the deformed WZ model. 
We consider that the character of the divergent nature 
and non-renormalizability of the SNJL model is unchanged by the deformation.
This model has no superpotential at this stage, and by applying the method of 
SUSY auxiliary fields of composites, the model will get (anti)holomorphic superpotentials.
The parity symmetry of spatial inversion is maintained in the form of the Lagrange function 
in the ordinary ${\cal N}=1$ case~[30,31], and the deformation explicitly breaks the symmetry.
The global gauge symmetry of the theory is chosen as $U(1)$, and $\Phi_{+}$ and $\Phi_{-}$ are oppositely charged.
According to Eqs. (20) and (21), the symmetrization of products of (anti)chiral superfields themselves 
has been taken in the K\"{a}hler potential given above.
In the ordinary ${\cal N}=1$ SUSY theory, ${\cal L}^{(1)}_{I}$ was prepared for the dynamical generation of 
a Dirac mass, while ${\cal L}^{(2)}_{I}$ is suitable for obtaining 
left-handed and right-handed Majorana mass terms~[30,31].  
Historically, the SNJL model was first introduced to investigate the phenomenon of 
dynamical chiral symmetry breaking in SUSY field theory~[26,27],
and applied to the top quark condensation of electroweak symmetry breaking of the minimal SUSY Standard Model~[28].
An $SU(N_{c})$ SNJL model was used to describe phenomena of phase transitions in the early universe~[29].
Investigations on dynamical chiral symmetry breaking and (color)superconductivity of SUSY condensed matter systems
were done by (25) at the case of the ordinary ${\cal N}=1$ superspace with a SUSY-breaking mass~[30,31].

Our model has following (pseudo)symmetries.
The global flavor symmetry $SU(N)_{L}\times SU(N)_{R}$ will be introduced 
in the theory by the following definitions:
\begin{eqnarray}
\Phi_{+} \to e^{-i\Gamma_{+}}\Phi_{+}, \quad \bar{\Phi}_{+} \to \bar{\Phi}_{+}e^{i\Gamma_{+}}, \quad 
\Phi_{-} \to e^{-i\Gamma_{-}}\Phi_{+}, \quad \bar{\Phi}_{-} \to \bar{\Phi}_{-}e^{i\Gamma_{-}}, \quad 
\Gamma_{\pm} \equiv \sum^{N^{2}-1}_{I=1}\Gamma^{I}_{\pm}T_{I},
\end{eqnarray}
where, $\Phi_{+}$ ( $\Phi_{-}$ ) belongs to the (anti)fundamental representation $N$ ( $\bar{N}$ ) of $SU(N)$,
and all of $\Gamma^{I}_{\pm}$ are real $c$-numbers.
Thus, the normalization factor $T(f)$ and the second-order Casimir invariant $C(f)$ 
of the Hermitian generators $T_{I}$ of $SU(N)$ ( $I=1, \cdots, N^{2}-1$ ) satisfy
\begin{eqnarray}
{\rm tr}T_{I}T_{J} = T(f)\delta_{IJ} = \frac{1}{2}\delta_{IJ}, \quad 
C(f)\delta_{ij} = \sum^{N^{2}-1}_{I=1}(T_{I}T_{I})_{ij} = \frac{N^{2}-1}{2N}\delta_{ij}, \quad
T(f) = \frac{N}{N^{2}-1}C(f).
\end{eqnarray}
It should be mention that, $(e^{-i\Gamma_{\pm}}\Phi_{\pm})^{\dagger}\ne \bar{\Phi}_{\pm}e^{i\Gamma_{\pm}}$
in general. If the symmetry of the unitary group are gauged, the discussion becomes more involved
because of the deformation.
Since the BCS-NJL mechanism breaks a global symmetry, 
the definitions of symmetries we concern here are enough for our purpose. 
The Lagrangian has the following global Abelian symmetries:
\begin{eqnarray}
U(1)_{V}: &\quad& \Phi_{+} \to e^{i\alpha_{v}}\Phi_{+}, \quad \Phi_{-} \to e^{-i\alpha_{v}}\Phi_{-}, \quad
\bar{\Phi}_{+} \to e^{-i\alpha_{v}}\bar{\Phi}_{+}, \quad \bar{\Phi}_{-} \to e^{i\alpha_{v}}\bar{\Phi}_{-}, \\
U(1)_{A}: &\quad& \Phi_{+} \to e^{i\alpha_{a}}\Phi_{+}, \quad \Phi_{-} \to e^{i\alpha_{a}}\Phi_{-}, \quad
\bar{\Phi}_{+} \to e^{-i\alpha_{a}}\bar{\Phi}_{+}, \quad \bar{\Phi}_{-} \to e^{-i\alpha_{a}}\bar{\Phi}_{-}, 
\end{eqnarray} 
where, $\alpha_{v}$ and $\alpha_{a}$ are real.
On the other hand, the $R$-symmetry,
\begin{eqnarray}
& & U(1)_{R}: \theta \to e^{i\alpha_{r}}\theta, \quad \bar{\theta} \to e^{-i\alpha_{r}}\bar{\theta}, \quad \phi_{\pm} \to e^{2in\alpha_{r}}\phi_{\pm}, \quad \psi_{\pm} \to e^{2i(n-\frac{1}{2})\alpha_{r}}\psi_{\pm}, \quad
F_{\pm} \to e^{2i(n-1)\alpha_{r}}F_{\pm}, 
\end{eqnarray}
was broken by the deformation. The deformation parameters $C^{\alpha\beta}$ carry 
a non-vanishing $U(1)_{R}$-charge:
\begin{eqnarray}
C^{\alpha\beta} &\to& e^{2i\alpha_{r}}C^{\alpha\beta}. 
\end{eqnarray}
In fact, $C^{\alpha\beta}$ act as $R$-symmetry-breaking parameters.

Next, we examine the component field expression of the Lagrangian (25).
In Refs.~[15,16], component field expressions for ${\cal N}=1/2$ generic chiral sigma models 
on the deformed superspace were obtained: It is given by a sum of an undeformed part 
and a deformed part, and the latter is given as an infinite-order power series of $\det C$ 
with including infinite-order partial derivatives of a K\"{a}hler potential taken with respect to scalar fields.
We follow the results of Ref.~[16], and add the following extra terms of the first-order in $\det C$ to the undeformed model:
\begin{eqnarray}
{\cal L}_{def}(C) &\equiv& -\det C\sum_{i,j=\pm}F_{i}F_{j}\Bigl\{ \sum_{k=\pm}\frac{1}{2!}\frac{\partial^{3}K(\bar{\phi}_{\pm}, \phi_{\pm})}{\partial\phi_{i}\partial\phi_{j}\partial\bar{\phi}_{k}}\Box\bar{\phi}_{k}
+ \sum_{k,l=\pm}\frac{1}{2!}\frac{\partial^{4}K(\bar{\phi}_{\pm},\phi_{\pm})}{\partial\phi_{i}\partial\phi_{j}\partial\bar{\phi}_{k}\partial\bar{\phi}_{l}}\partial_{\mu}\bar{\phi}_{k}\partial^{\mu}\bar{\phi}_{l} \Bigr\}.  
\end{eqnarray}
In our case, the deformed part of the Lagrangian is given as a function of metrics, connections and curvature
tensors of the K\"{a}hler manifold.
With the well-known procedure of the theory of K\"{a}hler manifold, the undeformed part of the Lagrangian (25),
${\cal L}(C=0)$, will be obtained as follows:
\begin{eqnarray}
{\cal L}(C) &=& {\cal L}(C=0) + {\cal L}_{def}(C),  \\
{\cal L}(C=0) &=& \sum_{i,j=\pm}g_{ij^{*}} \bigl( \bar{F}_{j}F_{i} -\partial_{\mu}\bar{\phi}_{j}\partial^{\mu}\phi_{i} -i\bar{\psi}_{j}\bar{\sigma}^{\mu}\partial_{\mu}\psi_{i} \bigr)
+\sum_{i,j,k=\pm}\Bigl[ g_{ij^{*},k^{*}}(-\frac{1}{2}F_{i}\bar{\psi}_{j}\bar{\psi}_{k})+g_{ij^{*},k}(-\frac{1}{2}\bar{F}_{j}\psi_{i}\psi_{k}) \Bigr]   \nonumber \\
& & +\sum_{i,j,k=\pm}g_{ij^{*},k}(-i\bar{\psi}_{j}\bar{\sigma}^{\mu}\psi_{k}\partial_{\mu}\phi_{i}) + \sum_{i,j,k,l=\pm}g_{ij^{*},kl^{*}}\frac{1}{4}(\psi_{i}\psi_{k}\bar{\psi}_{j}\bar{\psi}_{l}).
\end{eqnarray}
The K\"{a}hler metric of our model is given in the following matrix form:
\begin{eqnarray}
\hat{g} &\equiv& \left(
\begin{array}{cc}
g_{++^{*}} & g_{-+^{*}} \\
g_{+-^{*}} & g_{--^{*}}
\end{array}
\right) 
= \left( 
\begin{array}{cc}
1+\frac{G_{1}}{N}\bar{\phi}_{-}\phi_{-}+\frac{G_{2}}{N}\bar{\phi}_{+}\phi_{+}, & 
\frac{G_{1}}{N}\bar{\phi}_{-}\phi_{+}+\frac{G_{2}}{N}\bar{\phi}_{+}\phi_{-} \\
\frac{G_{1}}{N}\bar{\phi}_{+}\phi_{-}+\frac{G_{2}}{N}\bar{\phi}_{-}\phi_{+}, &
1+\frac{G_{1}}{N}\bar{\phi}_{+}\phi_{+}+\frac{G_{2}}{N}\bar{\phi}_{-}\phi_{-},
\end{array}
\right),
\end{eqnarray}
and the notations of the definitions $g_{ij^{*},k}=\partial g_{ij^{*}}/\partial\phi_{k}$, etc., 
have been used in Eq. (37).
In the K\"{a}hler metric given above, the Hermiticity may be lost, $(\hat{g})^{\dagger}\ne \hat{g}$
because $(\phi_{\pm})^{\dagger}\ne \bar{\phi}_{\pm}$ in general.
Obviously, the deformed Lagrangian (25) still has the invariance under the K\"{a}hler transformation:
\begin{eqnarray}
K(\bar{\Phi}_{\pm},\Phi_{\pm}) &\to& K(\bar{\Phi}_{\pm},\Phi_{\pm}) + {\cal F}(\Phi_{\pm}) + \bar{\cal F}(\Phi_{\pm}),
\end{eqnarray}
where, ${\cal F}$ ( $\bar{\cal F}$ ) is an arbitrarily function of $\Phi_{\pm}$ ( $\bar{\Phi}_{\pm}$ ).
Now, the Euler equations for the auxiliary fields are found to be
\begin{eqnarray}
0 = \frac{\partial {\cal L}(C)}{\partial F_{i}} 
&=& \sum_{j=\pm}\Bigg[ g_{ij^{*}}\bar{F}_{j} -\frac{1}{2}\sum_{k=\pm}g_{ij^{*},k^{*}}\bar{\psi}_{j}\bar{\psi}_{k} -2\det CF_{j}\Bigl(\sum_{k=\pm}\frac{1}{2!}g_{jk^{*},i}\Box\bar{\phi}_{k}+\sum_{k,l=\pm}\frac{1}{2!}g_{jk^{*},il^{*}}\partial_{\mu}\bar{\phi}_{k}\partial^{\mu}\bar{\phi}_{l}\Bigr)\Biggl],  \\
0 = \frac{\partial {\cal L}(C)}{\partial \bar{F}_{i}} 
&=& \sum_{j=\pm}\Bigl[g_{ji^{*}}F_{j} -\frac{1}{2}\sum_{k=\pm}g_{ji^{*},k^{*}}\psi_{j}\psi_{k}\Bigr],
\end{eqnarray}
( $i=\pm$ ). From these results, one finds
\begin{eqnarray}
\left(
\begin{array}{c} 
F_{+} \\
F_{-}
\end{array}
\right) &=& \frac{1}{2}\hat{g}^{-1}\sum_{k=\pm}\frac{\partial \hat{g}}{\partial\bar{\phi}_{k}}
\left(
\begin{array}{c}
\psi_{+}\psi_{k} \\
\psi_{-}\psi_{k}
\end{array}
\right),   \\
\left(
\begin{array}{c}
\bar{F}_{+} \\
\bar{F}_{-}
\end{array}
\right) &=& \frac{1}{2}(\hat{g}^{T})^{-1}\sum_{k=\pm}\frac{\partial \hat{g}^{T}}{\partial\bar{\phi}_{k}}
\left(
\begin{array}{c}
\bar{\psi}_{+}\bar{\psi}_{k} \\
\bar{\psi}_{-}\bar{\psi}_{k}
\end{array}
\right) + (\hat{g}^{T})^{-1}\det C\Bigg[\sum_{k=\pm}\left(
\begin{array}{cc}
g_{+k^{*},+} & g_{-k^{*},+} \\
g_{+k^{*},-} & g_{-k^{*},-}
\end{array}
\right)\Box\bar{\phi}_{k}  \nonumber \\
& & + \sum_{k,l=\pm}\left( 
\begin{array}{cc}
g_{+k^{*},+l^{*}} & g_{-k^{*},+l^{*}} \\
g_{+k^{*},-l^{*}} & g_{-k^{*},-l^{*}}
\end{array}
\right)\partial_{\mu}\bar{\phi}_{k}\partial^{\mu}\bar{\phi}_{l}
\Bigg]\frac{1}{2}\hat{g}^{-1}\sum_{k=\pm}\frac{\partial \hat{g}}{\partial\bar{\phi}_{k}}
\left(
\begin{array}{c}
\psi_{+}\psi_{k} \\
\psi_{-}\psi_{k}
\end{array}
\right).
\end{eqnarray}
We find that the auxiliary fields $F_{\pm},\bar{F}_{\pm}$ will be completely eliminated 
through these Euler equations in our model without any ambiguities. 
It is found that $\det C$ enters into four-fermion interaction terms under quite unusual forms
through the elimination of all the auxiliary fields $F_{\pm}$ and $\bar{F}_{\pm}$.
Thus, the quantum dynamics of the ${\cal N}=1/2$ deformed SNJL model might not resemble to the ordinary case,
though it is difficult for us to estimate the strength of the interaction coming from the deformation
in the form obtained above.

\subsection{SUSY Auxiliary Fields of Composites}

In this subsection, we examine the method of SUSY auxiliary fields of composites on the deformed superspace.
In the Lagrangian given in the previous subsection,
we have chosen the form of the K\"{a}hler potential suitable to generate holomorphic and antiholomorphic
superpotentials of SUSY composites.
Our model Lagrangian (25) will be rewritten in the following form through 
the method of SUSY auxiliary fields~[27-31]:
\begin{eqnarray}
{\cal L}(C) &=& {\cal L}_{0}+{\cal L}_{H}+{\cal L}_{1}(C)+{\cal L}_{\bar{1}}+{\cal L}_{2}(C)+{\cal L}_{\bar{2}},
\end{eqnarray}
where,
\begin{eqnarray}
{\cal L}_{0} &\equiv& \Bigl[\bar{\Phi}_{+}\star\Phi_{+}+\bar{\Phi}_{-}\star\Phi_{-}\Bigr]_{\theta\theta\bar{\theta}\bar{\theta}},  \quad {\cal L}_{H} \equiv \Bigl[ \frac{N}{G_{1}}\bar{H}_{1}\star H_{1} + \frac{4N}{G_{2}}\bar{H}_{2}\star H_{2} \Bigr]_{\theta\theta\bar{\theta}\bar{\theta}}, \nonumber \\
{\cal L}_{1}(C) &\equiv& \Bigl[ S_{1}\star\bigl( \frac{N}{G_{1}}H_{1}-\Phi_{+}\star\Phi_{-} \bigr)|_{sym} \Bigr]_{\theta\theta}, \quad {\cal L}_{\bar{1}} \equiv \Bigl[ \bar{S}_{1}\star\bigl( \frac{N}{G_{1}}\bar{H}_{1}-\bar{\Phi}_{+}\star\bar{\Phi}_{-} \bigr)|_{sym} \Bigr]_{\bar{\theta}\bar{\theta}}, \nonumber \\
{\cal L}_{2}(C) &\equiv& \Bigl[ S_{2}\star\bigl( \frac{4N}{G_{2}}H_{2}-\Phi_{+}\star\Phi_{+}-\Phi_{-}\star\Phi_{-} \bigr)|_{sym} \Bigr]_{\theta\theta}, \quad {\cal L}_{\bar{2}} \equiv \Bigl[ \bar{S}_{2}\star\bigl( \frac{4N}{G_{2}}\bar{H}_{2}-\bar{\Phi}_{+}\star\bar{\Phi}_{+}-\bar{\Phi}_{-}\star\bar{\Phi}_{-} \bigr)|_{sym} \Bigr]_{\bar{\theta}\bar{\theta}}.
\end{eqnarray}
Here, the method of SUSY auxiliary fields has been applied under taking the symmetrized star products.
Only ${\cal L}_{1}(C)$ and ${\cal L}_{2}(C)$ have contributions coming from the deformation.
Our model has superpotentials with constraints:
$S_{1}$, $\bar{S}_{1}$, $S_{2}$ and $\bar{S}_{2}$ were introduced as Lagrange multiplier multiplets 
for the purpose to enforce
\begin{eqnarray}
H_{1} &=& \frac{G_{1}}{N}\Phi_{+}\star\Phi_{-}|_{sym}, \quad \bar{H}_{1} = \frac{G_{1}}{N}\bar{\Phi}_{+}\star\bar{\Phi}_{-}|_{sym}, \nonumber \\
H_{2} &=& \frac{G_{2}}{4N}(\Phi_{+}\star\Phi_{+}+\Phi_{-}\star\Phi_{-}), \quad \bar{H}_{2} = \frac{G_{2}}{4N}(\bar{\Phi}_{+}\star\bar{\Phi}_{+}+\bar{\Phi}_{-}\star\bar{\Phi}_{-}).   
\end{eqnarray}
$H_{i}$ and $\bar{H}_{i}$ ( $i=1,2$ ) are regarded as SUSY composites.
Because the Hermiticity was lost in our theory, 
$(S_{i})^{\dagger}\ne \bar{S}_{i}$ and $(H_{i})^{\dagger}\ne \bar{H}_{i}$ in general.
Let us confirm the equivalence of Lagrange functions (25) and (44) under the symmetrized star products.
To convert the expression of the Lagrangian (44) in the point-product form, 
we take into account $Q^{2}(Q^{2}\Phi_{\pm})=0$. 
Because $\det C(Q^{2}S_{i})(Q^{2}H_{i})$ coming from $S_{i}\star H_{i}|_{sym}$
do not contribute to the component form of the holomorphic superpotential of the Lagrangian
through $\theta\theta$-integration, we drop these terms. 
After performing integrations of $\theta$ by parts,
${\cal L}_{1}(C)$, ${\cal L}_{\bar{1}}$, ${\cal L}_{2}(C)$ and ${\cal L}_{\bar{2}}$ will be rewritten
in the forms of ordinary point-products:
\begin{eqnarray}
{\cal L}_{1}(C) &=& \Bigg[ \frac{N}{G_{1}}S_{1}H_{1}-S_{1}\Phi_{+}\Phi_{-}+ \det C S_{1}(Q^{2}\Phi_{+})(Q^{2}\Phi_{-})\Bigg]_{\theta\theta}, \nonumber \\
{\cal L}_{2}(C) &=& \Bigg[ \frac{4N}{G_{2}} S_{2}H_{2} - S_{2}(\Phi_{+}\Phi_{+}+\Phi_{-}\Phi_{-})
+\det C S_{2}\{ (Q^{2}\Phi_{+})^{2}+(Q^{2}\Phi_{-})^{2} \}  \Bigg]_{\theta\theta},    \nonumber \\
{\cal L}_{\bar{1}} &=& \Bigl[ \bar{S}_{1} \Bigl(\frac{N}{G_{1}}\bar{H}_{1}-\bar{\Phi}_{+}\bar{\Phi}_{-}\Bigr) \Bigr]_{\bar{\theta}\bar{\theta}}, \quad 
{\cal L}_{\bar{2}} = \Bigl[ \bar{S}_{2} \Bigl(\frac{4N}{G_{2}}\bar{H}_{2}-\bar{\Phi}_{+}\bar{\Phi}_{+} -\bar{\Phi}_{-}\bar{\Phi}_{-} \Bigr) \Bigr]_{\bar{\theta}\bar{\theta}}.
\end{eqnarray}
In this expression, it is clear for us that the constraints given in (46) are satisfied 
under taking the symmetrized star products.
After $S_{i}$ and $\bar{S}_{i}$ were eliminated, (44) recovers (25).
The component field expression of the Lagrangian (44) is found to be:
\begin{eqnarray}
{\cal L}_{0} &=& \bar{F}_{+}F_{+}-\partial_{\mu}\bar{\phi}_{+}\partial^{\mu}\phi_{+} -i\bar{\psi}_{+}\bar{\sigma}^{\mu}\partial_{\mu}\psi_{+} + \bar{F}_{-}F_{-}-\partial_{\mu}\bar{\phi}_{-}\partial^{\mu}\phi_{-}-i\bar{\psi}_{-}\bar{\sigma}^{\mu}\partial_{\mu}\psi_{-}, \\
{\cal L}_{H} &=& \frac{N}{G_{1}}\Bigl( \bar{F}_{H_{1}}F_{H_{1}}-\partial_{\mu}\bar{\phi}_{H_{1}}\partial^{\mu}\phi_{H_{1}}-i\bar{\psi}_{H_{1}}\bar{\sigma}^{\mu}\partial_{\mu}\psi_{H_{1}} \Bigr) + \frac{4N}{G_{2}}\Bigl( \bar{F}_{H_{2}}F_{H_{2}}-\partial_{\mu}\bar{\phi}_{H_{2}}\partial^{\mu}\phi_{H_{2}}-i\bar{\psi}_{H_{2}}\bar{\sigma}^{\mu}\partial_{\mu}\psi_{H_{2}} \Bigr), \\
{\cal L}_{1}(C) &=& \frac{N}{G_{1}}\Bigl( \phi_{S_{1}}F_{H_{1}}+F_{S_{1}}\phi_{H_{1}}-\psi_{S_{1}}\psi_{H_{1}}\Bigr) \nonumber \\
& & -\Bigl( F_{S_{1}}\phi_{+}\phi_{-}+\phi_{S_{1}}F_{+}\phi_{-}+\phi_{S_{1}}\phi_{+}F_{-}
-\phi_{S_{1}}\psi_{+}\psi_{-}-\phi_{+}\psi_{S_{1}}\psi_{-}-\phi_{-}\psi_{S_{1}}\psi_{+} \Bigr) +\det CF_{S_{1}}F_{+}F_{-},  \\
{\cal L}_{\bar{1}} &=& ({\cal L}_{1}(C=0))^{\dagger}, \\
{\cal L}_{2}(C) &=& \frac{4N}{G_{2}}\Bigl( \phi_{S_{2}}F_{H_{2}}+F_{S_{2}}\phi_{H_{2}}-\psi_{S_{2}}\psi_{H_{2}}\Bigr) -\Bigl( F_{S_{2}}(\phi^{2}_{+}+\phi^{2}_{-})   \nonumber \\ 
& & +\phi_{S_{2}}(2F_{+}\phi_{+}+2F_{-}\phi_{-}-\psi_{+}\psi_{+}-\psi_{-}\psi_{-} )-2\psi_{S_{2}}(\phi_{+}\psi_{+}+\phi_{-}\psi_{-}) \Bigr) +\det CF_{S_{2}}(F^{2}_{+}+F^{2}_{-}),  \\
{\cal L}_{\bar{2}} &=& ({\cal L}_{2}(C=0))^{\dagger}.  
\end{eqnarray}
We observe ${\cal L}(C)={\cal L}(C=0)+\det C\{ F_{S_{1}}F_{+}F_{-} + F_{S_{2}}(F^{2}_{+}+F^{2}_{-})\}$, 
a sum of ordinary Lagrangian with deformed part, the result of Ref.~[16] holds in the case of 
the model with SUSY auxiliary fields of composites.
If $\phi_{S_{1}}$ and $\bar{\phi}_{S_{1}}$ ( Dirac mass ) or $\phi_{S_{2}}$ and $\bar{\phi}_{S_{2}}$ ( Majorana mass ) obtain finite VEVs, 
superfields $\Phi_{\pm}$ and $\bar{\Phi}_{\pm}$ become massive. 
The Euler equations for the auxiliary fields of chiral multiplets are found to be
\begin{eqnarray}
0 &=& \frac{\partial {\cal L}(C)}{\partial F_{+}} 
= \bar{F}_{+}-\phi_{S_{1}}\phi_{-}-2\phi_{S_{2}}\phi_{+}+\det C (F_{S_{1}}F_{-}+2F_{S_{2}}F_{+}),  \quad 0 = \frac{\partial {\cal L}(C)}{\partial \bar{F}_{+}}
= F_{+}-\bar{\phi}_{S_{1}}\bar{\phi}_{-}-2\bar{\phi}_{S_{2}}\bar{\phi}_{+},   \nonumber \\
0 &=& \frac{\partial {\cal L}(C)}{\partial F_{-}}
= \bar{F}_{-}-\phi_{S_{1}}\phi_{+}-2\phi_{S_{2}}\phi_{-}+\det C (F_{S_{1}}F_{+}+2F_{S_{2}}F_{-}),  \quad 0 = \frac{\partial {\cal L}(C)}{\partial \bar{F}_{-}}  
= F_{-}-\bar{\phi}_{S_{1}}\bar{\phi}_{+}-2\bar{\phi}_{S_{2}}\bar{\phi}_{-},   \nonumber \\
0 &=& \frac{\partial {\cal L}(C)}{\partial F_{S_{1}}} 
= \frac{N}{G_{1}}\phi_{H_{1}}-\phi_{+}\phi_{-}+\det CF_{+}F_{-}, \quad 0 = \frac{\partial {\cal L}(C)}{\partial \bar{F}_{S_{1}}} 
= \frac{N}{G_{1}}\bar{\phi}_{H_{1}}-\bar{\phi}_{+}\bar{\phi}_{-},  \nonumber \\
0 &=& \frac{\partial {\cal L}(C)}{\partial F_{S_{2}}}
= \frac{4N}{G_{2}}\phi_{H_{2}}-\phi^{2}_{+}-\phi^{2}_{-}+\det C(F^{2}_{+}+F^{2}_{-}),  \quad 0 = \frac{\partial {\cal L}(C)}{\partial \bar{F}_{S_{2}}} 
= \frac{4N}{G_{2}}\bar{\phi}_{H_{2}}-\bar{\phi}^{2}_{+}-\bar{\phi}^{2}_{-},  \nonumber \\
0 &=& \frac{\partial {\cal L}(C)}{\partial F_{H_{1}}} 
= \frac{N}{G_{1}}(\bar{F}_{H_{1}}+\phi_{S_{1}}),  \quad 0 = \frac{\partial {\cal L}(C)}{\partial \bar{F}_{H_{1}}} 
= \frac{N}{G_{1}}(F_{H_{1}}+\bar{\phi}_{S_{1}}),    \nonumber \\
0 &=& \frac{\partial {\cal L}(C)}{\partial F_{H_{2}}}
= \frac{4N}{G_{2}}(\bar{F}_{H_{2}}+\phi_{S_{2}}),  \quad 0 = \frac{\partial {\cal L}(C)}{\partial \bar{F}_{H_{2}}}
= \frac{4N}{G_{2}}(F_{H_{2}}+\bar{\phi}_{S_{2}}).   
\end{eqnarray}
Hence, $F_{\pm}$, $\bar{F}_{\pm}$, $F_{H_{i}}$, and $\bar{F}_{H_{i}}$ ( $i=1,2$ ) 
can be eliminated from ${\cal L}(C)$ in the component field form.
After the elimination of auxiliary fields through the Euler equations, one finds 
\begin{eqnarray}
\phi_{H_{1}} &=& \frac{G_{1}}{N}\Bigl( \phi_{+}\phi_{-}-\det C(\bar{\phi}_{S_{1}}\bar{\phi}_{-}+2\bar{\phi}_{S_{2}}\bar{\phi}_{+})(\bar{\phi}_{S_{1}}\bar{\phi}_{+}+2\bar{\phi}_{S_{2}}\bar{\phi}_{-}) \Bigr),  \quad \bar{\phi}_{H_{1}} = \frac{G_{1}}{N}\bar{\phi}_{+}\bar{\phi}_{-}, \nonumber \\
\phi_{H_{2}} &=& \frac{G_{2}}{4N}\Bigl( \phi^{2}_{+}+\phi^{2}_{-}-\det C\bigl\{ (\bar{\phi}_{S_{1}}\bar{\phi}_{-}+2\bar{\phi}_{S_{2}}\bar{\phi}_{+})^{2}+(\bar{\phi}_{S_{1}}\bar{\phi}_{+}+2\bar{\phi}_{S_{2}}\bar{\phi}_{-})^{2}\bigr\} \Bigr), \quad \bar{\phi}_{H_{2}} = \frac{G_{2}}{4N}(\bar{\phi}^{2}_{+}+\bar{\phi}^{2}_{-}).
\end{eqnarray}
From these expressions, again we have found that the expressions of the constraints of 
SUSY composites given in (46) are satisfied in terms of the scalar fields.
The Lagrange multiplier multiplets $S_{i}$ and $\bar{S}_{i}$ ( $i=1,2$ ) work well to keep the
constraints of SUSY composites, and guarantee the equivalence of the Lagrange functions (25) and (44).
It should be remarked that, $(\phi_{H_{i}})^{\dagger}\ne\bar{\phi}_{H_{i}}$ ( $i=1,2$ ): 
The Hermiticity was explicitly lost in the component field level of the SUSY composites at $\det C\ne 0$.
Now, we find the tree-level potential $V_{tree}$ of this model (44):
\begin{eqnarray}
V_{tree} &\equiv& -\frac{N}{G_{1}}\Bigl( \bar{F}_{H_{1}}F_{H_{1}}+\phi_{S_{1}}F_{H_{1}}+\bar{\phi}_{S_{1}}\bar{F}_{H_{1}}+ \phi_{H_{1}}F_{S_{1}}+\bar{\phi}_{H_{1}}\bar{F}_{S_{1}} \Bigr) \nonumber \\
& & -\frac{4N}{G_{2}}\Bigl( \bar{F}_{H_{2}}F_{H_{2}}+\phi_{S_{2}}F_{H_{2}}+\bar{\phi}_{S_{2}}\bar{F}_{H_{2}}+ \phi_{H_{2}}F_{S_{2}}+\bar{\phi}_{H_{2}}\bar{F}_{S_{2}} \Bigr).
\end{eqnarray}
Thus, from 
\begin{eqnarray}
0 = \frac{\partial V_{tree}}{\partial \phi_{H_{1}}} = -\frac{N}{G_{1}}F_{S_{1}},  \quad 
0 = \frac{\partial V_{tree}}{\partial \bar{\phi}_{H_{1}}} = -\frac{N}{G_{1}}\bar{F}_{S_{1}}, \quad
0 = \frac{\partial V_{tree}}{\partial \phi_{H_{2}}} = -\frac{4N}{G_{2}}F_{S_{2}},  \quad 
0 = \frac{\partial V_{tree}}{\partial \bar{\phi}_{H_{2}}} = -\frac{4N}{G_{2}}\bar{F}_{S_{2}},  \nonumber \\
0 = \frac{\partial V_{tree}}{\partial F_{S_{1}}} = -\frac{N}{G_{1}}\phi_{H_{1}},  \quad 
0 = \frac{\partial V_{tree}}{\partial \bar{F}_{S_{1}}} = -\frac{N}{G_{1}}\bar{\phi}_{H_{1}},  \quad
0 = \frac{\partial V_{tree}}{\partial F_{S_{2}}} = -\frac{4N}{G_{2}}\phi_{H_{2}},  \quad 
0 = \frac{\partial V_{tree}}{\partial \bar{F}_{S_{2}}} = -\frac{4N}{G_{2}}\bar{\phi}_{H_{2}}, 
\end{eqnarray}
one finds that the global minimum of $V_{tree}$ locates at $F_{S_{1}}=\bar{F}_{S_{1}}=F_{S_{2}}=\bar{F}_{S_{2}}=0$.
Through the Euler equations of $F_{H_{1}}$, $\bar{F}_{H_{1}}$, $F_{H_{2}}$ and $\bar{F}_{H_{2}}$ given in (54), we find
\begin{eqnarray}
V_{tree}&=& -\frac{N}{G_{1}}(\phi_{H_{1}}F_{S_{1}}+\bar{\phi}_{H_{1}}\bar{F}_{S_{1}}-\bar{\phi}_{S_{1}}\phi_{S_{1}})
-\frac{4N}{G_{2}}(\phi_{H_{2}}F_{S_{2}}+\bar{\phi}_{H_{2}}\bar{F}_{S_{2}}-\bar{\phi}_{S_{2}}\phi_{S_{2}})= \frac{N}{G_{1}}\bar{\phi}_{S_{1}}\phi_{S_{1}}+\frac{4N}{G_{2}}\bar{\phi}_{S_{2}}\phi_{S_{2}}.
\end{eqnarray}
Here, we have used the conditions (57).
Thus, from the Euler equations of $\phi_{S_{1}}$, $\bar{\phi}_{S_{1}}$, $\phi_{S_{2}}$ and $\bar{\phi}_{S_{2}}$, 
one finds $V_{tree}$ has a global minimum at
\begin{eqnarray}
\phi_{S_{1}} = \bar{\phi}_{S_{1}} = \phi_{S_{2}} = \bar{\phi}_{S_{2}} = 0.
\end{eqnarray}
Obviously, the vacuum energy vanishes at the global minimum.
It should be noticed that $V_{tree}$ at the global minimum is Hermite in spite of the non-Hermitian character of ${\cal L}(C)$.
At the classical tree-level, the symmetries are not broken spontaneously in our model.
From the discussion given above, one recognizes that $F_{S_{1}}$ and $\bar{F}_{S_{1}}$ 
may have a special role to keep the symmetries of the theory unbroken.
We will go beyond the classical tree-level analysis, 
examine what kind of modifications will happen under the nonperturbatibe quantum dynamics in the SNJL model,
and whether there is a crucial change in the effective action for the investigation of 
the dynamical mass generation in the BCS-NJL mechanism.

\subsection{The One-loop Effective Action}

In this subsection, we evaluate the effective action at one-loop level~[33,34],
derive and solve the gap equation for the examination whether the dynamical mass generation
takes place or not in the ${\cal N}=1/2$ SNJL model.
There are several methods for this purpose. 
For example, we can employ the method of SUSY auxiliary fields with the steepest descent technique
for a self-consistent evaluation of composite condensates with superfield or 
component field formalisms of the Lagrangian.
Another choice for us is the method of Schwinger-Dyson equation based on a diagrammatic consideration
under keeping star products or after converting them into ordinary point-products.
For introducing the method of Cornwall-Jackiw-Tomboulis effective action of composites~[25],
we have to extend it to the ${\cal N}=1/2$ SUSY case, and it may have external bilocal sources in superspace 
as conjugates of composites with the operations of star products, thus it will demand us
to do a complicated work.  
We choose the method of effective action of the Lagrangian with SUSY auxiliary fields 
in superspace, and perform the large-$N$ expansion ( a kind of the steepest descent technique ) in this subsection. 
It is clear from the Lagrangian (44), the contribution coming from the loop-expansion ( except the tree level ) 
does not includes $\phi_{H_{i}}$, $\bar{\phi}_{H_{i}}$, $F_{H_{i}}$ and $\bar{F}_{H_{i}}$.
After $\Phi_{\pm}$ and $\bar{\Phi}_{\pm}$ were integrated out, 
the contribution of loop-expansion to the effective action gives quantum corrections
expressed as a function of $\phi_{S_{i}}$, $\bar{\phi}_{S_{i}}$, $F_{S_{i}}$ and $\bar{F}_{S_{i}}$.

Hereafter, we consider the case $G_{2}=0$. The effective action $\Gamma$ of the theory is given by
\begin{eqnarray}
\Gamma &\equiv& -i\ln\int [d\Phi_{+}][d\bar{\Phi}_{+}][d\Phi_{-}][d\bar{\Phi}_{-}][dH_{1}][d\bar{H}_{1}][dS_{1}][d\bar{S}_{1}]\exp\Bigl[ i\int d^{4}y {\cal L}_{T} \Bigr],  \\
{\cal L}_{T} &\equiv& {\cal L}(C) 
+ \bigl[m\Phi_{+}\star\Phi_{-}|_{sym}\bigr]_{\theta\theta}
+ \bigl[\bar{m}\bar{\Phi}_{+}\star\bar{\Phi}_{-}|_{sym}\bigr]_{\bar{\theta}\bar{\theta}} = {\cal L}(C) 
+ \bigl[ m\Phi_{+}\Phi_{-} \bigr]_{\theta\theta}
+ \bigl[ \bar{m}\bar{\Phi}_{+}\bar{\Phi}_{-} \bigr]_{\bar{\theta}\bar{\theta}}.
\end{eqnarray}
Here, ${\cal L}(C)$ takes the form of Eq. (44).
We have introduced a holomorphic and an anti-holomorphic mass terms ( correspond to a Dirac mass term for fermions in the ordinary ${\cal N}=1$ SUSY case )
to make the effective action infrared well-defined~[11,35].
In general, both $m$ and $\bar{m}$ are matrices in the flavor space.
When the mass terms take diagonal and flavor-independent forms in the flavor space,
they break $U(N)_{L}\times U(N)_{R}$ into the diagonal subgroup $U(N)_{L+R}$. 
In the case of deformed superspace, we can take $m^{\dagger}\ne\bar{m}$ in general.
The result of component field expression of ${\cal L}(C)$ given in (48)-(53) should be kept in our mind,
and calculation will be performed in the similar way to the ordinary ${\cal N}=1$ case~[32,33,34].
All of symmetrized star products are reducible to ordinary point-products.
First, we convert all of the symmetrized star products in our Lagrangian to ordinary point products, 
and then put the model to the formalism of effective action in superspace.
When $\Phi_{\pm}$ are chiral superfields, $Q^{2}\Phi_{\pm}$ are also chiral superfields.
Thus, the Feynman rules of the deformed superspace are obtained exactly the same way with the ordinary case,
if we treat $Q^{2}\Phi_{\pm}$ as independent chiral superfields~[11,12]. 
We can utilize this fact for the diagrammatic consideration for our calculation.
The evaluation of the effective action should be performed under the superfield formalism
because we consider a nonperturbative problem, and we have to sum up a subset of Feynman diagrams of the theory.
In the case we concern in this paper, it seems difficult ( or, hard ) to achieve this purpose under the component field formalism.

Now, we take a steepest descent approximation in the path integrations of the auxiliary fields $H_{1}$, $\bar{H}_{1}$, $S_{1}$ and $\bar{S}_{1}$ in (60), 
regard all of the SUSY auxiliary fields as classical constant $c$-numbers.
We expand ${\cal L}_{T}$ in terms of quantum fluctuating fields $\Phi_{\pm}$, $\bar{\Phi}_{\pm}$ 
around the origin $\Phi_{\pm}=\bar{\Phi}_{\pm}=0$:
\begin{eqnarray}
{\cal L}_{T} &=& {\cal L}_{T}[\Phi_{\pm}=\bar{\Phi}_{\pm}=0] + {\cal L}_{(2)} + \cdots, \\
{\cal L}_{(2)} &\equiv& \frac{1}{2}\int d^{2}\theta d^{2}\bar{\theta}( \bar{\Phi}_{+}, \bar{\Phi}_{-}, \Phi_{+}, \Phi_{-} ){\cal M}
\left(
\begin{array}{c}
\Phi_{+} \\
\Phi_{-} \\
\bar{\Phi}_{+} \\
\bar{\Phi}_{-}
\end{array}
\right).
\end{eqnarray}
The rules for functional derivatives of the variation of chiral and antichiral superfields
\begin{eqnarray}
\frac{\delta\Phi_{\pm}(z')}{\delta \Phi_{\pm}(z)} = -\frac{1}{4}\overline{D}^{2}\delta^{8}(z-z'), \quad 
\frac{\delta\bar{\Phi}_{\pm}(z')}{\delta \bar{\Phi}_{\pm}(z)} = -\frac{1}{4}D^{2}\delta^{8}(z-z')
\end{eqnarray}
( $z, z'=(y,\theta,\bar{\theta})$ ) are still satisfied also in our case. Therefore,
\begin{eqnarray}
{\cal M} &\equiv& \left(
\begin{array}{cc}
\frac{\overrightarrow{\delta}}{\delta\bar{\Phi}_{\pm}}{\cal L}_{T}[\Phi_{\pm},\bar{\Phi}_{\pm}]\frac{\overleftarrow{\delta}}{\Phi_{\pm}}|_{\Phi_{\pm}=\bar{\Phi}_{\pm}=0}
&
\frac{\overrightarrow{\delta}}{\delta\bar{\Phi}_{\pm}}{\cal L}_{T}[\Phi_{\pm},\bar{\Phi}_{\pm}]\frac{\overleftarrow{\delta}}{\bar{\Phi}_{\pm}}|_{\Phi_{\pm}=\bar{\Phi}_{\pm}=0}
\\
\frac{\overrightarrow{\delta}}{\delta\Phi_{\pm}}{\cal L}_{T}[\Phi_{\pm},\bar{\Phi}_{\pm}]\frac{\overleftarrow{\delta}}{\Phi_{\pm}}|_{\Phi_{\pm}=\bar{\Phi}_{\pm}=0}
&
\frac{\overrightarrow{\delta}}{\delta\Phi_{\pm}}{\cal L}_{T}[\Phi_{\pm},\bar{\Phi}_{\pm}]\frac{\overleftarrow{\delta}}{\bar{\Phi}_{\pm}}|_{\Phi_{\pm}=\bar{\Phi}_{\pm}=0}
\end{array}
\right) \nonumber \\
&=& \left(
\begin{array}{cc}
-\frac{D^{2}\overline{D}^{2}}{16}\otimes\sigma_{0}  
&   
(-\frac{D^{2}}{4})(\bar{m}+\bar{S}_{1})\otimes\sigma_{1}  
\\
(-\frac{\overline{D}^{2}}{4})\bigl[m+S_{1}-\det C(Q^{2}S_{1})Q^{2}\bigr]\otimes\sigma_{1} 
& 
-\frac{\overline{D}^{2}D^{2}}{16}\otimes\sigma_{0}
\end{array}
\right)\delta^{8}(z-z'),
\end{eqnarray}
where, we have used the equivalent relations $\delta^{2}(\bar{\theta})=-D^{2}/4\Box$ and 
$\delta^{2}(\theta)=-\overline{D}^{2}/4\Box$ inside the integration of the action functional.  
The relations $D^{2}\overline{D}^{2}D^{2}=16\Box D^{2}$, 
$\overline{D}^{2}D^{2}\overline{D}^{2}=16\Box\overline{D}^{2}$ were also used.
If we consider the case $G_{2}\ne 0$ with $G_{1}=0$, the off-diagonal part of ${\cal M}$ will be replaced as
$-D^{2}[\bar{m}+\bar{S}_{1}]\otimes\sigma_{1}/4\to -D^{2}[\bar{m}\otimes\sigma_{1}+\bar{S}_{2}\otimes 1]/4$ and
$-\overline{D}^{2}[m+S_{1}-\det C(Q^{2}S_{1})Q^{2}]\otimes\sigma_{1}/4\to -\overline{D}^{2}[m\otimes\sigma_{1}+(S_{2}-\det C(Q^{2}S_{2})Q^{2})\otimes 1]/4$.
Except these replacements, the formulae we will obtain hereafter are essentially the same for the $G_{2}\ne 0$ case.
The component field expressions of the SUSY auxiliary fields are given by
\begin{eqnarray}
S_{1} = \phi_{S_{1}}+\theta^{2}F_{S_{1}}, \quad \bar{S}_{1} = \bar{\phi}_{S_{1}}+\bar{\theta}^{2}\bar{F}_{S_{1}}.
\end{eqnarray}
It should be noted that $\theta^{2}$ ( $\bar{\theta}^{2}$ ) acts as a (anti)chiral superfield.
After putting these expressions to (60) and performing the Berezinian integration, one gets
\begin{eqnarray}
\Gamma &=& \int d^{4}y{\cal L}_{T}[\Phi_{\pm}=\bar{\Phi}_{\pm}=0] + \tilde{\Gamma}, \\
\tilde{\Gamma} &\equiv& -i\ln\int[d\Phi_{+}][d\bar{\Phi}_{+}][d\Phi_{-}][d\bar{\Phi}_{-}] \exp\Bigl[ i 
\int d^{4}y{\cal L}_{(2)} +\cdots \Bigr] = \frac{i}{2}\ln{\rm Det}{\cal M} + \cdots.
\end{eqnarray}
Hence, the effective action at the one-loop level is found to be
\begin{eqnarray}
\Gamma &=& N\int d^{4}y\Bigg[ 
\frac{1}{G_{1}}\bar{H}_{1}H_{1} + \frac{1}{G_{1}}S_{1}H_{1}\delta^{2}(\bar{\theta})+\frac{1}{G_{1}}\bar{S}_{1}\bar{H}_{1}\delta^{2}(\theta)\Bigg]_{\theta\theta\bar{\theta}\bar{\theta}} + \lim_{z'\to z}\frac{i}{2}N\int d^{8}z{\rm tr}(\ln{\cal M})\delta^{8}(z-z'). 
\end{eqnarray} 
At $S_{1}=\bar{S}_{1}=0$, $\tilde{\Gamma}$ identically vanishes because it corresponds to a tadpole graph
in the ${\cal N}=1$ case. The effect of deformation contained in the operator $[\det C(Q^{2}S_{1})]Q^{2}$ 
has to act in the inside of loop integrals. 
The evaluation of $\tilde{\Gamma}$ is just to extract the $D$-term contribution after
some manipulations of $Q_{\alpha}$, $D_{\alpha}$ and $\overline{D}_{\dot{\alpha}}$.
We use several identities in the integrals of the loop expansion:
\begin{eqnarray}
\frac{1}{16}D^{2}\overline{D}^{2}\delta^{2}(\theta-\theta')\delta^{2}(\bar{\theta}-\bar{\theta}')|_{\theta=\theta',\bar{\theta}=\bar{\theta}'} &=& 1, \\
\frac{1}{16}Q^{2}\overline{D}^{2}\delta^{2}(\theta-\theta')\delta^{2}(\bar{\theta}-\bar{\theta}')|_{\theta=\theta',\bar{\theta}=\bar{\theta}'} &=& 1, \\
Q^{2}\frac{1}{16}D^{2}\overline{D}^{2}\delta^{2}(\theta-\theta')\delta^{2}(\bar{\theta}-\bar{\theta}')|_{\theta=\theta',\bar{\theta}=\bar{\theta}'} &=& -4\bar{\theta}\bar{\theta}\Box.
\end{eqnarray}
${\cal M}$ will be divided into the following form:
\begin{eqnarray}
{\cal M} &\equiv& {\cal M}_{0} - \Sigma,  \nonumber \\
{\cal M}_{0} &\equiv& \left(
\begin{array}{cc}
\frac{D^{2}\overline{D}^{2}}{16}  &   0  \\
0 & \frac{\overline{D}^{2}D^{2}}{16}
\end{array}
\right), \quad 
{\cal M}^{-1}_{0} = \frac{1}{\Box^{2}}\left(
\begin{array}{cc}
\frac{D^{2}\overline{D}^{2}}{16}  &   0  \\
0 & \frac{\overline{D}^{2}D^{2}}{16}
\end{array}
\right),  \nonumber \\
\Sigma &\equiv& \left(
\begin{array}{cc}
0  &  -(-\frac{D^{2}}{4})(\bar{m}+\bar{S}_{1})\otimes\sigma_{1}  \\
-(-\frac{\overline{D}^{2}}{4})\bigl[ m+ S_{1} -\det C(Q^{2}S_{1})Q^{2} \bigr]\otimes\sigma_{1} & 0
\end{array}
\right).
\end{eqnarray}
Thus, 
\begin{eqnarray}
\tilde{\Gamma} &=& \frac{i}{2}{\rm Tr} \Bigl[ \ln \bigl( {\cal M}_{0} - \Sigma \bigr) \Bigr] = \frac{i}{2}{\rm Tr}\ln{\cal M}_{0} + \frac{i}{2}{\rm Tr}\ln\Bigl(1-{\cal M}^{-1}_{0}\Sigma \Bigr) = \frac{i}{2}{\rm Tr}\ln\Bigl(1-{\cal M}^{-1}_{0}\Sigma \Bigr).
\end{eqnarray}
We have dropped $\frac{i}{2}{\rm Tr}\ln{\cal M}_{0}$ because it does not contribute to the integral of 
$\tilde{\Gamma}$. 
Because the inverse matrix of chiral projectors obeys the relations
${\cal M}^{-1}_{0}{\cal M}^{-1}_{0}=\frac{1}{\Box}{\cal M}^{-1}_{0}$, $[{\cal M}^{-1}_{0}, \Sigma]=0$,
and all of the operators in ${\cal M}^{-1}_{0}$ and $\Sigma$ are bosonic,
the expression given above will be converted as
\begin{eqnarray} 
{\rm Tr}\ln(1-{\cal M}^{-1}_{0}\Sigma) =
-{\rm Tr}\Bigl\{  {\cal M}^{-1}_{0}\Sigma 
+ \frac{1}{2}{\cal M}^{-1}_{0}\Sigma{\cal M}^{-1}_{0}\Sigma
+ \frac{1}{3}{\cal M}^{-1}_{0}\Sigma{\cal M}^{-1}_{0}\Sigma{\cal M}^{-1}_{0}\Sigma + \cdots  \Bigr\} = {\rm Tr}\Bigg[ \Bigl( \ln\bigl[ 1-\frac{\Sigma}{\Box} \bigr] \Bigr) \Box{\cal M}^{-1}_{0} \Biggr].
\end{eqnarray}
Therefore, one gets $\tilde{\Gamma}$ in the following form:
\begin{eqnarray}
\tilde{\Gamma} &=& \lim_{z'\to z}\frac{i}{2}N{\rm tr}\int d^{4}yd^{2}\theta d^{2}\bar{\theta} 
\ln\Bigl[ 1-\frac{1}{\Box}(\bar{m}+\bar{S}_{1})\bigl\{ m+ S_{1} -\det C(Q^{2}S_{1})Q^{2} \bigr\} \Bigr]\frac{D^{2}\overline{D}^{2}}{16\Box}\delta^{8}(z-z')    \nonumber \\
&=& -\frac{i}{2}N{\rm tr}\int d^{4}yd^{2}\theta d^{2}\bar{\theta}\int \frac{d^{4}p}{(2\pi)^{4}}
\frac{1}{p^{2}}\ln\Bigl( 1+\frac{1}{p^{2}}(\bar{m}+\bar{S}_{1})\bigl\{ m+ S_{1} -\det C(Q^{2}S_{1})Q^{2} \bigr\} \Bigr) \frac{D^{2}\overline{D}^{2}}{16}\delta^{2}(\theta-\theta')\delta^{2}(\bar{\theta}-\bar{\theta}').    \nonumber \\
& & 
\end{eqnarray}
The effective potential is defined by
\begin{eqnarray}
V \equiv -\frac{\Gamma}{\int d^{4}y}, \qquad 
\tilde{V} \equiv -\frac{\tilde{\Gamma}}{\int d^{4}y} .
\end{eqnarray}

It is a quite difficult ( hard ) problem to evaluate $\tilde{\Gamma}$ or $\tilde{V}$ exactly.
Let us first consider the small deformation limit $|\det C|\to 0$, and expand $\tilde{V}$ as 
a power series of $\det C$:
\begin{eqnarray}
\tilde{V} &=& \frac{i}{2}N{\rm tr}\int d^{2}\theta d^{2}\bar{\theta} 
\int\frac{d^{4}p}{(2\pi)^{4}}\frac{1}{p^{2}}\Bigg[ 
\ln\{ p^{2}+(\bar{m}+\bar{S}_{1})(m+ S_{1}) \}      \nonumber \\
& & - \sum^{\infty}_{n=1}\frac{(\det C)^{n}}{n}\Bigl(
\frac{(\bar{m}+\bar{S}_{1})(Q^{2}S_{1})}{p^{2}+(\bar{m}+\bar{S}_{1})(m+ S_{1})}Q^{2}\Bigr)^{n} \Bigg]
\frac{D^{2}\overline{D}^{2}}{16}\delta^{2}(\theta-\theta')\delta^{2}(\bar{\theta}-\bar{\theta}')  \nonumber \\
&=& \tilde{V}(\det C=0) + (\det C)\tilde{V}^{(1)} + (\det C)^{2}\tilde{V}^{(2)} + \cdots.
\end{eqnarray}
It is a kind of perturbative analysis on the effect of the interaction coming from the deformation.
( Note that it is not a perturbative expansion in $G_{1}$. )
The zero-th order in $\det C$ corresponds to the undeformed case~[33,34]:
\begin{eqnarray}
\tilde{V}(\det C=0) 
&=& \frac{i}{2}N{\rm tr}\int d^{2}\theta d^{2}\bar{\theta}\int \frac{d^{4}p}{(2\pi)^{4}}\frac{1}{p^{2}}\ln\Bigl( 1+\frac{1}{p^{2}}(\bar{m}+\bar{S}_{1})(m+ S_{1}) \Bigr)  \nonumber \\
&=& -\frac{N}{16\pi^{2}}\Bigl[ 
\{\Lambda^{2}+(\bar{m}+\bar{S}_{1})(m+ S_{1})\}\ln\{\Lambda^{2}+(\bar{m}+\bar{S}_{1})(m+ S_{1})\}     \nonumber \\
& & \quad -(\bar{m}+\bar{S}_{1})(m+ S_{1})\ln(\bar{m}+\bar{S}_{1})(m+ S_{1}) \Bigr]_{\theta\theta\bar{\theta}\bar{\theta}},    
\end{eqnarray}
where, a Euclidean four-dimensional cutoff $\Lambda^{2}$ was introduced for the regularization of the integral.
This term is proportional to $\bar{F}_{S_{1}}F_{S_{1}}$ 
and it corresponds to a radiatively induced K\"{a}hler potential
for $S_{1}$ and $\bar{S}_{1}$ which is not included in the classical ${\cal L}(C)$ or the tree-level action $\Gamma-\tilde{\Gamma}$.
In the expansion $\tilde{V}=\tilde{V}(\det C=0)+(\det C)\tilde{V}^{(1)}+\cdots$ given above, 
only for the first term $\tilde{V}(\det C=0)$ we can utilize the method of K\"{a}hler geometry 
to obtain the component field expression of it.
$\tilde{V}^{(1)}$, the contribution for the first order in $\det C$, is found to be
\begin{eqnarray}
\tilde{V}^{(1)} &=& \frac{i}{2}N{\rm tr}\int d^{2}\theta d^{2}\bar{\theta}\int \frac{d^{4}p}{(2\pi)^{4}}
\frac{1}{p^{2}}\frac{(\bar{m}+\bar{S}_{1})(-1)(Q^{2}S_{1})}{p^{2}+(\bar{m}+\bar{S}_{1})(m+ S_{1})}
Q^{2}\frac{D^{2}\overline{D}^{2}}{16}\delta^{2}(\theta-\theta')\delta^{2}(\bar{\theta}-\bar{\theta}')  \nonumber \\
&=& -\frac{N}{4\pi^{2}}\Bigl[
(\bar{m}+\bar{S}_{1})^{2}(m+S_{1})(Q^{2}S_{1})\bar{\theta}\bar{\theta}\ln\bigl(1+\frac{\Lambda^{2}}{(\bar{m}+\bar{S}_{1})(m+S_{1})} \bigr)
\Bigr]_{\theta\theta\bar{\theta}\bar{\theta}}.
\end{eqnarray}
This integral gives a term proportional to $F^{2}_{S_{1}}$, 
a new term generated by the quantum radiative correction.
In the case of the WZ model, the one-loop correction of the first-order in $\det C$
gives the term $[\Phi Q^{2}\Phi]_{\theta\theta}$ proportional to $F^{2}$~[11,12]. 
The second-order corrections in our case seems to have a complicated structure if we convert it to the
component field expression, because there are insertions of the derivative operator $Q^{2}$ 
between the functions of superfields $S_{1}$ and $\bar{S}_{1}$.
However, because $Q^{2}Q^{2}D^{2}\overline{D}^{2}\delta^{2}(\theta)\delta^{2}(\bar{\theta})=0$, one finds
\begin{eqnarray}
\tilde{V}^{(2)} &=& \frac{i}{2}N{\rm tr}\int d^{2}\theta d^{2}\bar{\theta}\int \frac{d^{4}p}{(2\pi)^{4}}\frac{1}{p^{2}} \frac{(\bar{m}+\bar{S}_{1})(Q^{2}S_{1})}{p^{2}+(\bar{m}+\bar{S}_{1})(m+ S_{1})}
Q^{2}
\frac{(\bar{m}+\bar{S}_{1})(Q^{2}S_{1})}{p^{2}+(\bar{m}+\bar{S}_{1})(m+ S_{1})}
Q^{2}\frac{D^{2}\overline{D}^{2}}{16}\delta^{2}(\theta-\theta')\delta^{2}(\bar{\theta}-\bar{\theta}')  \nonumber \\
&=& \frac{N}{8\pi^{2}}\Bigl[ \frac{\Lambda^{4}(\bar{m}+\bar{S}_{1})^{2}(Q^{2}S_{1})^{3}\bar{\theta}\bar{\theta}}{(m+S_{1})\{\Lambda^{2}+(\bar{m}+\bar{S}_{1})(m+S_{1})\}^{2}}  \Bigr]_{\theta\theta\bar{\theta}\bar{\theta}}.
\end{eqnarray}
This integral gives the term proportional to $F^{4}_{S_{1}}$.
We speculate these new exotic terms will arise infinitely, and $V$ takes a form which has 
several similar features with the component field expression of 
the classical-level of a generic sigma model obtained by Rittov and Sannino~[16]. 
By gathering the results of the analysis of power-series in $\det C$, 
the one-loop effective potential of the leading order of the large-$N$ expansion is obtained 
in the following component field form:
\begin{eqnarray}
\frac{V}{N} &=& -\frac{1}{G_{1}}\Bigl( \bar{F}_{H_{1}}F_{H_{1}}+\phi_{S_{1}}F_{H_{1}}+\bar{\phi}_{S_{1}}\bar{F}_{H_{1}}+ \phi_{H_{1}}F_{S_{1}}+\bar{\phi}_{H_{1}}\bar{F}_{S_{1}} \Bigr) \nonumber \\
& & + \bar{F}_{S_{1}}F_{S_{1}}f^{(0)}(\phi_{S_{1}},\bar{\phi}_{S_{1}})+ (\det C)F^{2}_{S_{1}}f^{(1)}(\phi_{S_{1}},\bar{\phi}_{S_{1}})+ (\det C)^{2}F^{4}_{S_{1}}f^{(2)}(\phi_{S_{1}},\bar{\phi}_{S_{1}})+ \cdots,  
\end{eqnarray}
where, we have used the following functions for writing our formula compactly:
\begin{eqnarray}
f^{(0)}(\phi_{S_{1}},\bar{\phi}_{S_{1}}) &\equiv& -\frac{1}{16\pi^{2}}\Bigl[ \ln\bigl( 1 + \frac{\Lambda^{2}}{(\bar{m}+\bar{\phi}_{S_{1}})(m+\phi_{S_{1}})} \bigr)
-\frac{\Lambda^{2}}{\Lambda^{2}+(\bar{m}+\bar{\phi}_{S_{1}})(m+\phi_{S_{1}})}  \Bigr], \\
f^{(1)}(\phi_{S_{1}},\bar{\phi}_{S_{1}}) &\equiv& -\frac{1}{4\pi^{2}} 
(\bar{m}+\bar{\phi}_{S_{1}})^{2} \Bigl[ \ln \bigl( 1+\frac{\Lambda^{2}}{(\bar{m}+\bar{\phi}_{S_{1}})(m+\phi_{S_{1}})} \bigr)
-\frac{\Lambda^{2}}{\Lambda^{2}+(\bar{m}+\bar{\phi}_{S_{1}})(m+\phi_{S_{1}})}\Bigr], \\
f^{(2)}(\phi_{S_{1}},\bar{\phi}_{S_{1}}) &\equiv& -\frac{1}{8\pi^{2}}\Bigl(\frac{\bar{m}+\bar{\phi}_{S_{1}}}{m+\phi_{S_{1}}}\Bigr)^{2}\frac{\Lambda^{4}[\Lambda^{2}+3(\bar{m}+\bar{\phi}_{S_{1}})(m+\phi_{S_{1}})]}{[\Lambda^{2}+(\bar{m}+\bar{\phi}_{S_{1}})(m+\phi_{S_{1}})]^{3}}.
\end{eqnarray}
$f^{(0)}(\phi_{S_{1}},\bar{\phi}_{S_{1}})$ is proportional to the K\"{a}hler metric which can be obtained from the K\"{a}hler potential given in the final result of Eq. (79).
$f^{(0)}$ and $f^{(1)}$ diverge logarithmically, while the function $f^{(2)}$ for the second-order contribution $\tilde{V}^{(2)}$ remains finite at $\Lambda^{2}\to\infty$: $\lim_{\Lambda^{2}\to\infty}f^{(2)}=-(\bar{m}+\bar{\phi}_{S_{1}})^{2}/[8\pi^{2}(m+\phi_{S_{1}})^{2}]$.
We observe the operator $Q^{2}$ acts on $G(p^{2})$, where,
\begin{eqnarray}
G(p^{2}) \equiv \frac{(\bar{m}+\bar{S}_{1})(Q^{2}S_{1})}{p^{2}+(\bar{m}+\bar{S}_{1})(m+S_{1})},
\end{eqnarray} 
in the expansion of $\tilde{V}$, and this causes the modifications of the momentum dependence of $G(p^{2})$, $Q^{2}G(p^{2})$, $Q^{2}(Q^{2}G(p^{2}))$, ..., as follows:
\begin{eqnarray}
G(p^{2}) \sim {\cal O}(p^{-2}), \quad Q^{2}G(p^{2}) \sim {\cal O}(p^{-4}), \quad Q^{2}(Q^{2}G(p^{2})) \sim {\cal O}(p^{-6}), \quad \cdots.
\end{eqnarray}
Therefore, the insertions of the operator $Q^{2}$ in $\tilde{V}$ given in Eq. (78) always lower the orders of divergences.
Hence, the one-loop contribution at the large $\Lambda^{2}$ region, namely $\Lambda^{2}\gg |(\bar{m}+\bar{\phi}_{S_{1}})(m+\phi_{S_{1}})|$, behaves as follows:
\begin{eqnarray}
\tilde{V} &\sim& -\frac{1}{16\pi^{2}}\Bigl[\bar{F}_{S_{1}}F_{S_{1}}+4F^{2}_{S_{1}}\det C(\bar{m}+\bar{\phi}_{S_{1}})^{2} \Bigr]\ln\frac{\Lambda^{2}}{(\bar{m}+\bar{\phi}_{S_{1}})(m+\phi_{S_{1}})}.
\end{eqnarray}
The divergent behavior of $\tilde{V}$ at $\Lambda^{2}\to\infty$ is determined by the zeroth- and the first-order terms of $\det C$, 
the contributions higher than ${\cal O}(\det C)$ become small, and the ultraviolet divergence of $\tilde{V}$ is exactly logarithmic. 
Again, as same as the procedure performed in the previous subsection,
we take variations with respect to the components of the classical SUSY auxiliary fields for obtaining their Euler-Lagrange equations 
( coupled nonlinear gap equations ) to determine the vacuum state:
\begin{eqnarray}
0 = \frac{1}{N}\frac{\partial V}{\partial F_{H_{1}}} &=& -\frac{1}{G_{1}}(\bar{F}_{H_{1}}+\phi_{S_{1}}), \quad
0 = \frac{1}{N}\frac{\partial V}{\partial \bar{F}_{H_{1}}} = -\frac{1}{G_{1}}(F_{H_{1}}+\bar{\phi}_{S_{1}}),   \nonumber \\
0 = \frac{1}{N}\frac{\partial V}{\partial \phi_{H_{1}}} &=& -\frac{1}{G_{1}}F_{S_{1}},  \quad
0 = \frac{1}{N}\frac{\partial V}{\partial \bar{\phi}_{H_{1}}} = -\frac{1}{G_{1}}\bar{F}_{S_{1}},   \nonumber \\
0 = \frac{1}{N}\frac{\partial V}{\partial F_{S_{1}}} &=& -\frac{1}{G_{1}}\phi_{H_{1}} + \bar{F}_{S_{1}}f^{(0)}(\phi_{S_{1}},\bar{\phi}_{S_{1}}) + 2(\det C)F_{S_{1}}f^{(1)}(\phi_{S_{1}},\bar{\phi}_{S_{1}})
+ 4(\det C)^{2}F^{3}_{S_{1}}f^{(2)}(\phi_{S_{1}},\bar{\phi}_{S_{1}})  + \cdots,   \nonumber \\
0 = \frac{1}{N}\frac{\partial V}{\partial \bar{F}_{S_{1}}} &=& -\frac{1}{G_{1}}\bar{\phi}_{H_{1}} 
+ F_{S_{1}}f^{(0)}(\phi_{S_{1}},\bar{\phi}_{S_{1}}) + \cdots,   \nonumber \\
0 = \frac{1}{N}\frac{\partial V}{\partial \phi_{S_{1}}} &=& -\frac{1}{G_{1}}F_{H_{1}} + \bar{F}_{S_{1}}F_{S_{1}}\frac{\partial}{\partial\phi_{S_{1}}}f^{(0)}(\phi_{S_{1}},\bar{\phi}_{S_{1}})  \nonumber \\
& & + (\det C)F^{2}_{S_{1}}\frac{\partial}{\partial\phi_{S_{1}}}f^{(1)}(\phi_{S_{1}},\bar{\phi}_{S_{1}})
+ (\det C)^{2}F^{4}_{S_{1}}\frac{\partial}{\partial\phi_{S_{1}}}f^{(2)}(\phi_{S_{1}},\bar{\phi}_{S_{1}}) + \cdots,   \nonumber \\
0 = \frac{1}{N}\frac{\partial V}{\partial \bar{\phi}_{S_{1}}} &=& -\frac{1}{G_{1}}\bar{F}_{H_{1}} 
+ \bar{F}_{S_{1}}F_{S_{1}}\frac{\partial}{\partial\bar{\phi}_{S_{1}}}f^{(0)}(\phi_{S_{1}},\bar{\phi}_{S_{1}})   \nonumber \\
& & + (\det C)F^{2}_{S_{1}}\frac{\partial}{\partial\bar{\phi}_{S_{1}}}f^{(1)}(\phi_{S_{1}},\bar{\phi}_{S_{1}}) 
+ (\det C)^{2}F^{4}_{S_{1}}\frac{\partial}{\partial\bar{\phi}_{S_{1}}}f^{(2)}(\phi_{S_{1}},\bar{\phi}_{S_{1}}) + \cdots.  
\end{eqnarray}
Now we find that the VEV of $F_{S_{1}}$ ( $\bar{F}_{S_{1}}$ ) is still zero because 
$\phi_{H_{1}}$ ( $\bar{\phi}_{H_{1}}$ ) only couples with $F_{S_{1}}$ ( $\bar{F}_{S_{1}}$ ).
Each terms in $V$ are given as factorized forms of products of powers of $F_{S_{1}}$, $\bar{F}_{S_{1}}$ and 
$f^{(n)}(\phi_{S_{1}},\bar{\phi}_{S_{1}})$ ( namely, like $(\det C)^{l}F^{m}_{S_{1}}\bar{F}^{n}_{S_{1}}f^{(n,m)}$, $l,m,n=0,1,2,\cdots$ ), 
thus the vacuum energy vanishes at the origin: The global minimum of $V_{tree}$ is not changed.
From this fact, the composite field $\phi_{S_{1}}$ ( $\bar{\phi}_{S_{1}}$ ) cannot obtain a non-zero VEV.
The expansion of $V$ given in Eq. (82) shows an asymmetric form in terms of the auxiliary fields $F_{S_{1}}$ and $\bar{F}_{S_{1}}$,
and the coefficient functions $f^{(1)}$ and $f^{(2)}$ are also asymmetric with respect to $\phi_{S_{1}}$ and $\bar{\phi}_{S_{1}}$, while $f^{(0)}$ is symmetric.
The quantum fluctuations ( massive ) of the composite fields might be affected by the asymmetry of $V$ 
because the curvature tensor ( corresponds to the mass tensor of the collective fields ) of $V$ at the global minimum might have a skew structure,
while it is not clear whether the effect of deformation enhances or reduces the amplitudes of fluctuations around the origin of $V$.

We believe that the analysis of power-series expansion in $\det C$ is useful for our qualitative understanding of $V$,
though the convergence of the series might be questionable. 
To go beyond the analysis of the power-series expansion of $\det C$, 
we employ the following approximation to evaluate (76).
The power series of $\det C$ of $\tilde{V}$ given in Eq. (78) is written symbolically as
\begin{eqnarray}
-{\rm tr}\Bigl[ (\det C)Q^{2}G + \frac{(\det C)^{2}}{2}Q^{2}GQ^{2}G + \cdots  \Bigr],
\end{eqnarray}
where, a cyclic permutation of $Q^{2}$ inside the trace has been taken.
Now, we employ the approximation where $Q^{2}$ acts only the nearest neighbor $G$ in the products $Q^{2}GQ^{2}G\cdots Q^{2}G$.
Therefore, each terms will be factorized as $Q^{2}GQ^{2}G\cdots Q^{2}G \to (Q^{2}G)^{n}$,
and then it becomes possible to sum up the infinite-order power series.
( If we take into account all of possible operations of $Q^{2}$ in $Q^{2}GQ^{2}G\cdots Q^{2}G$, integer factors determined from the numbers of possible insertions of $Q^{2}$ should be multiplied to $(\det C)^{n}(Q^{2}G)^{n}/n$, and then it becomes impossible to sum up the power series (90) in the form of a logarithmic function given below, Eq. (91). )
By employing this approximation, we obtain the following result:
\begin{eqnarray}
\tilde{V} &=& \frac{i}{2}N{\rm tr}\int d^{2}\theta d^{2}\bar{\theta} \int\frac{d^{4}p}{(2\pi)^{4}}\frac{1}{p^{2}} \nonumber \\
& & \times \ln \Bigl[ p^{2}+(\bar{m}+\bar{S}_{1})(m+ S_{1})     
+\det C\frac{(\bar{m}+\bar{S}_{1})^{2}(Q^{2}S_{1})^{2}}{[p^{2}+(\bar{m}+\bar{S}_{1})(m+ S_{1})]^{2}} \Bigr] \frac{D^{2}\overline{D}^{2}}{16}\delta^{2}(\theta-\theta')\delta^{2}(\bar{\theta}-\bar{\theta}').
\end{eqnarray}
Here, $\det C$ enters into $\tilde{V}$ in the "nonperturbative" way.
At $\det C=0$, this result recovers $\tilde{V}(\det C=0)$ of Eq. (79).
This result of the one-loop effective potential shows a skew structure,
i.e. the Hermiticity of $\tilde{V}$ was lost by the deformation.
The (anti)commutators presented in Sec. II guarantee that $\overline{Q}_{\dot{\alpha}}$
never appear in the theory, and thus still ${\cal N}=1/2$ is maintained in the expression of the effective potential given above. 
If we neglect the $p^{2}$-dependence of the coefficient function of $\det C$ inside the argument of the $\log$ function, 
Eq.(91) is evaluated into the following compact form:
\begin{eqnarray}
\tilde{V} &=& \frac{i}{2}N{\rm tr}\int d^{2}\theta d^{2}\bar{\theta} \int\frac{d^{4}p}{(2\pi)^{4}}\frac{1}{p^{2}} 
\ln \Bigl[ p^{2}+(\bar{m}+\bar{S}_{1})(m+ S_{1}) +\det C\frac{(Q^{2}S_{1})^{2}}{(m+ S_{1})^{2}} \Bigr]     \nonumber \\
&=& -\frac{N}{16\pi^{2}}\Bigg[ 
\Bigl\{\Lambda^{2}+(\bar{m}+\bar{S}_{1})(m+ S_{1})+\det C\frac{(Q^{2}S_{1})^{2}}{(m+ S_{1})^{2}}\Bigr\}\ln\Bigl\{\Lambda^{2}+(\bar{m}+\bar{S}_{1})(m+ S_{1})+\det C\frac{(Q^{2}S_{1})^{2}}{(m+ S_{1})^{2}}\Bigr\}     \nonumber \\
& & \quad -\Bigl\{(\bar{m}+\bar{S}_{1})(m+ S_{1})+\det C\frac{(Q^{2}S_{1})^{2}}{(m+ S_{1})^{2}}\Bigr\}\ln\Bigl\{(\bar{m}+\bar{S}_{1})(m+ S_{1})+\det C\frac{(Q^{2}S_{1})^{2}}{(m+ S_{1})^{2}}\Bigr\} \Bigg]_{\theta\theta\bar{\theta}\bar{\theta}}     \nonumber \\
&=& -\frac{N}{16\pi^{2}}\bar{F}_{S_{1}}F_{S_{1}}\Bigl[ \ln(1+\frac{\Lambda^{2}}{\alpha})-\{(\bar{m}+\bar{\phi}_{S_{1}})(m+\phi_{S_{1}})-2F^{2}_{S_{1}}\det C(m+\phi_{S_{1}})^{-2}\} \frac{\Lambda^{2}}{\alpha(\Lambda^{2}+\alpha)}\Bigr],  \\
\alpha &\equiv& (\bar{m}+\bar{\phi}_{{S}_{1}})(m+\phi_{S_{1}})+\frac{F^{2}_{S_{1}}\det C}{(m+ \phi_{S_{1}})^{2}}.
\end{eqnarray}
Here, we can check the correctness of our calculation by the examination of the mass dimension
of the result (92) by using $\phi_{S_{1}}\sim [{\rm mass}]^{1}$ and $F_{S_{1}}\sim [{\rm mass}]^{2}$.
From the same reason discussed in the analysis of the Euler-Lagrange equations (89) obtained by the power series expansion of $\det C$ given above,
we recognize that, there is no dynamical generation of mass in the SNJL model
on ${\cal N}=1/2$ deformed superspace: 
All of the VEVs of $\phi_{H_{1}}$, $\bar{\phi}_{H_{1}}$, $\phi_{S_{1}}$ and $\bar{\phi}_{S_{1}}$ are zero,
and the dynamical mass generation never occurs. 
The functional structure of our $V$ has some similarities with the component field expression of a generic sigma model on ${\cal N}=1/2$ non(anti)commutative superspace. 
In a generic sigma model, the component field expression is
always given as a power series of the auxiliary field $F$ and $\bar{F}$~[16].
The case we concern here is not the same with the generic sigma model, because our $V$ of (82) or (92) cannot be converted into
a form of a sigma model of the star products of chiral superfields,
though it is the case that $V$ will also be expanded in terms of $F_{S_{1}}$ and $\bar{F}_{S_{1}}$ ( an analytic function with respect to these fields ), 
and given in the forms where $\det C$ is always multiplied to $F_{S_{1}}$ directly ( i.e., like the form $(\det C)^{l}F^{n}_{S_{1}}$ ).
Due to this functional structure, 
the deformation effect vanishes at the origin $F_{S_{1}}=\bar{F}_{S_{1}}=0$ which gives the variationally determined vacuum of our model, 
hence $V$ restores the Hermiticity in the vacuum state of our model.
In the WZ model, the renormalization to the holomorphic superpotential take place, 
while the nonrenormalization theorem still be kept in the anti-holomorphic part.
One of the origins of the differences of the effects of quantum radiative corrections of 
the SNJL model compared with the WZ model is coming from the role of 
the classical SUSY auxiliary fields of composites in the one-loop effective action of the SNJL model. 
In the WZ model, the tree-level potential is given by classical background (anti)chiral superfields themselves.

It was known in literature~[26-31], in the ordinary ${\cal N}=1$ SNJL model, 
the vacuum stabilization by a dynamical mass generation can take place if 
${\cal N}=1$ SUSY is broken by adding a non-holomorphic SUSY-breaking mass term
$\Delta^{2}_{soft}(\bar{\phi}_{+}\phi_{+}+\bar{\phi}_{-}\phi_{-})$ for scalars.
This term gives a finite vacuum energy to the theory.
The situation we have considered here is different from that: 
The ${\cal N}=1$ SUSY was broken to ${\cal N}=1/2$ but the vacuum energy vanishes. 
The symmetric vacuum state of the theory is stable against the radiative correction,
no tachyonic mode, and thus the dynamical mass generation through the BCS-NJL mechanism of SUSY composites does not take place.
It is clear from our formulation, the inclusion of Majorana mass by setting $G_{2}\ne 0$ will also give the similar conclusion.

\section{Concluding Remarks}

In summary, we have examined the nonperturbative quantum dynamics of the SNJL model 
on ${\cal N}=1/2$ deformed superspace by evaluating the one-loop effective action 
in the leading-order of the large-$N$ expansion in the superfield formalism, 
which includes the infinite-order power series of $\det C$. 
We have understood the structure of the ${\cal N}=1/2$ SNJL model with the SUSY auxiliary fields,
not only by the superfield formalism, but also by the component field expression of it.
We have shown that the method of SUSY auxiliary fields of composites can be introduced
to the ${\cal N}=1/2$ SNJL model in the same way with the ${\cal N}=1$ ordinary case.
By using the one-loop effective action, we have shown that the nonperturbative quantum dynamics
gives infinite number of new exotic terms they do not appear in the ordinary undeformed case of SNJL model,
and the effect of the deformation gives distortion around the global minimum of the effective potential.
The global minimum of the theory at the tree level is unchanged by the quantum correction, and the dynamical mass generation never takes place:
These new terms are not relevant for giving the dynamical mass generation.
The broken Hermiticity is restored at the global minimum of the effective potential,
and the variationally determined ground state coincides with that of the ordinary ${\cal N}=1$ SNJL model.
It is an interesting fact that the ${\cal N}=1/2$ SNJL model will choose the vacuum of the ${\cal N}=1$ case,
thus the effect of the deformation vanishes at the vacuum state of our model:
The global minimum of the ${\cal N}=1/2$ SNJL model has "effectively" the ${\cal N}=1$ SUSY.

In this paper, we have introduced the ${\cal N}=1/2$ deformed SNJL model
with implementing the star product of the (anti)chiral superfields themselves,
and our calculation has been done naively by the usual method one-loop effective potential in the ${\cal N}=1$ case.
One of the important remaining issues of this work is that, 
whether the usual method of loop-expansion in the ordinary ${\cal N}=1$ case is mathematically satisfactorily rigorous also in
the deformed superspace or not.
In other words, it is an interesting problem for us to construct a mathematically rigorous treatment of the method of evaluation of an effective action
in the deformed superspace. 
This problem might be related to the theory of superanalysis of quantum superspace~[36].

It is also interesting for us on the examination on the possible cooperation
of the soft mass $\Delta^{2}_{soft}\theta^{2}\bar{\theta}^{2}\bar{\Phi}_{\pm}\Phi_{\pm}$ 
and the new interactions coming from the deformation of SNJL model in dynamical mass generation. 
The SNJL model with a SUSY-breaking mass spontaneously breaks the chiral symmetry,
though the inclusion of the SUSY-breaking mass will make our calculation more involved one than the results given in this paper.
Another problem is the examination on the quantum fluctuation effect
of the SUSY auxiliary fields of composites, or by taking into account the next-leading-order
of the loop-expansion to obtain propagators of the collective fields,
though these examination might not alter the global minimum of $V$ given in this work crucially.
One could observe an effect of the deformed superspace in quantum fluctuations due to the skew structure around the global minimum of the effective potential 
( from an examination on the curvature tensor = mass tensor of the collective modes ), 
though usually, quantum fluctuations of a theory will vanish at the large-$N$ limit $N\to\infty$:
The effect of the deformation of superspace would be suppressed by the (semi)classical nature of the theory.

The deformation of the superspace introduced by Eq. (1) was found in the analyses
of the supergravity and superstring theories.
Here, we wish to change the direction of our consideration.
One of the interesting issue for us is to apply our method to the investigation 
of a nonperturbative quantum dynamics of ${\cal N}=1$ four-dimensional superstring effective theory
on ${\cal N}=1/2$ non(anti)commutative superspace.
In that issue, the non(anti)commutation relations are required from the starting point.
It is well-known fact that, results of various compactification schemes of several superstring models
share generic features, and they are summarized in the following form, as a kind of
nonlinear sigma model~[37,38]:
\begin{eqnarray}
K_{string} &=& -\ln(S+\bar{S}) -3\ln(T+\bar{T}-\alpha'\bar{\Psi}\Psi) = -\ln(S+\bar{S}) +3\Bigl[ \frac{\alpha'\bar{\Psi}\Psi}{T+\bar{T}}
+\frac{1}{2}\bigl(\frac{\alpha'\bar{\Psi}\Psi}{T+\bar{T}}\bigr)^{2}+\cdots \Bigr],  \\
W_{string} &=& g\Psi\Psi\Psi.
\end{eqnarray} 
Here, $K_{string}$ is a K\"{a}hler potential, $W_{string}$ is a holomorphic superpotential, 
$S$ is a dilaton, $T$ is a moduli superfield, $\Psi$ is a chiral matter multiplet as an irreducible representation of a gauge group.
If the expansion of $K_{string}$ is truncated at the second-order of $\alpha'$,
the model has a similar structure of the SNJL model we have discussed in this paper.
Thus, the application of our method to the model given above
in the case of non(anti)commutative superspace is possible and an interesting issue.
Because the Hermiticity was lost in the deformed superspace, 
this fact might affect not only matter fields but also dilaton or moduli
under both classical and quantum levels, and an effective action of this model
might have a very skew structure, while the Hermiticity might be restored at the global minimum of this model. 
After integrate out $\Psi$ and $\bar{\Psi}$, an effective action will be given as a functional of
$S$, $\bar{S}$, $T$ and $\bar{T}$ ( possibly in an infinite-order power series of $\det C$
generated by a nonperturbative quantum dynamics ). 
Both dilaton and moduli fields are candidates of the inflaton.
If a gravitational interaction is taken into account under a suitable ( consistent ) way,
the theory of the model might give several exotic results or predictions
to the structure of the universe, inflation, and so forth,
and the predictions might relate to cosmological observations.

\end{document}